\documentstyle[12pt]{article}
\newcommand{\nc}{\newcommand}

\newcommand{\my}{\setcounter{equation}{0}}

\nc{\zm}[1]{\stackrel{o} {#1} }
\nc{\nm}[1]{\stackrel{n} {#1} }
\begin{document}
\begin{flushright}
MPI H-V19-1997\\
DPNU-97-22
\end{flushright}

\begin{center}
{\Large {\bf Zero Mode and Symmetry Breaking\\ 
 on the Light Front }}
\vspace{25pt}
\noindent

Sho Tsujimaru $\mathop{}^a$
\footnote{E-mail address: sho@mickey.mpi-hd.mpg.de} 
and 
Koichi Yamawaki $\mathop{}^{b}$
\footnote{E-mail address: yamawaki@eken.phys.nagoya-u.ac.jp}

$\mathop{}^{a}$ Max-Planck-Institut f\"ur Kernphysik, 
D-69029 Heidelberg\\
$\mathop{}^{b}$ Department of Physics,  Nagoya University, 
Nagoya 464-01, Japan
\end{center}
\vspace{15pt}
\begin{abstract} 
We study the zero mode and the spontaneous symmetry breaking 
on the light front (LF).
We use the discretized 
light-cone quantization (DLCQ) of Maskawa-Yamawaki 
to treat the zero mode in a clean separation
from all other modes. 
It is then shown that the Nambu-Goldstone (NG) phase can be 
realized on the trivial LF vacuum only when an 
explicit symmetry-breaking mass of the NG boson $m_{\pi}$  
is introduced. 
The NG-boson zero mode integrated over the 
LF must exhibit singular behavior $ \sim 1/m_{\pi}^2$
in the symmetric limit $m_{\pi}\rightarrow 0$, which 
implies that current conservation is
violated at zero mode, or equivalently the 
LF charge is not conserved even in the symmetric limit.  
We demonstrate this peculiarity in a concrete model, 
the linear sigma model,
where the role of zero-mode constraint is clarified. 
We further compare our result with the continuum theory. 
It is shown that in the continuum theory it is difficult 
to remove the zero mode which is not a single mode with 
measure zero but the accumulating 
point causing uncontrollable infrared singularity. 
A possible way out within the continuum theory
is also suggested based on the ``$\nu$ theory''.
We finally discuss another problem of the zero mode 
in the continuum theory, i.e., 
no-go theorem of Nakanishi-Yamawaki on the non-existence of
LF quantum field theory within the framework of Wightman
axioms, which remains to be a challenge for DLCQ, ``$\nu$ theory''
or any other framework of LF theory.
\end{abstract}
\begin{center}
\it {To appear in Physical Review D}. 
\end{center}
\newpage
\setlength{\baselineskip}{0.3in}
\section{Introduction}\par
\my

Recent revival of the light-front (LF) 
quantization \cite{Dira} aims at establishing a new formulation 
to study non-perturbative 
dynamics \cite{review,Wils}. 
A striking feature of the LF field theories is 
that the LF vacuum is simple, or even trivial \cite{KLS}.
The subtlety on this conclusion due to 
the so-called ``zero mode'' was first 
addressed back in 1976 by Maskawa and Yamawaki \cite{MY} 
who proposed the 
{\it discretized light-cone quantization} 
(DLCQ) \footnote{
The name ``light-cone quantization'' is actually confusing, 
since it is not on the 
light cone but on the light front which agrees with the former 
only in $1+1$ dimensions. However, 
here we simply follow the conventional naming of the majority of 
the literature. 
The DLCQ was also considered by Casher \cite{casher}
independently in a different context.  
}
to treat the zero mode in a clean separation from other modes. 
They found a
constraint equation for the zero mode 
({\it ``zero-mode constraint''}) 
through which the zero mode becomes dependent on other 
modes and can in principle be removed from the physical 
Fock space by solving the zero-mode constraint, thus 
{\it establishing the trivial 
vacuum in DLCQ.}

Based on the notion of this trivial vacuum,  
the first application of DLCQ to non-perturbative calculation  
was done in 1985 by Pauli and Brodsky \cite{PB} in the context of 
(1+1)-dimensional Yukawa model: 
The DLCQ  offers the promising prescription for obtaining 
the relativistic wave functions  
and the bound-state spectra in gauge theories. 
Their scheme has been applied to various models such as 
$\phi^4_{2}$ theory \cite{scalar}, Abelian \cite{qed2} as well as 
non-Abelian \cite{qcd2} gauge theories  in (1+1) dimensions and  
the models in  four 
dimensions \cite{TBP}. 
As far as the  two-dimensional 
models are concerned, 
reasonable correspondences with the known results have been obtained.
 
In spite of the success in two dimensions, 
there are a number of problems which must be 
solved to apply the same method to realistic models in four 
dimensions such as QCD. 
One of such problems is the long-standing zero-mode 
problem \cite{MY}. 
While the triviality of the LF vacuum in DLCQ can be 
achieved by solving out the zero mode from the physical
Fock space through the zero-mode constraint \cite{MY}, 
such a trivial vacuum would confront the usual 
picture of the complicated non-perturbative 
vacuum structure in the conventional 
equal-time quantization such as the confinement and the spontaneous 
symmetry breaking (SSB). Simplicity of the LF vacuum and states can 
in fact only
be realized at the sacrifice of simplicity of the operator side: The
only
operator responsible for such phenomena should 
be the zero mode whose constraint 
actually carries essential information of the complicated dynamics.
One might thus expect that explicit solution of the zero-mode 
constraint in DLCQ should give rise to the physics equivalent to the 
nontrivial vacuum structure in equal-time quantization,  
while preserving the trivial LF vacuum.
Actually, such an idea was carried out in the case of 
(1+1)-dimensional $\phi^4$ model \cite{HKSW,Rob,BPV} 
where it was 
argued 
that 
the solution of the zero-mode constraint 
might
lead to the SSB (of a discrete symmetry). 
However, the most outstanding feature of the SSB is the existence
of the Nambu-Goldstone (NG) boson associated with the continuous
symmetry breaking in four
dimensions.

In this paper, we elaborate on our previous paper \cite{KTY}
to examine how the NG boson in four dimensions
can be described  on the LF
in view of the zero mode in DLCQ. 
The main conclusion of  the previous 
work was that 
contrary to the naive expectation mentioned above,
solving the zero-mode 
constraint does not lead to the NG phase, 
unless we introduce an explicit 
symmetry breaking (mass of the NG boson).  
{\it The NG phase can only be
realized 
when the NG-boson zero mode integrated 
over the LF behaves as $\sim 1/m_{\pi}^2$ in the symmetry 
limit $m_{\pi}^2 \rightarrow 0$.} 
The most striking feature of its consequence is that the LF charge 
(zero mode of the local current) corresponding to the SSB
is {\it not} conserved even
in the symmetry limit $m_{\pi}^2 \rightarrow 0$.

Following Maskawa and Yamawaki \cite{MY} 
we formulate the canonical DLCQ 
{\' a} la Dirac for the scalar theory with
a periodic boundary condition. 
Then the zero mode can be treated separately from
other mode and be removed out of the physical Fock space
through the zero-mode constraint \cite{MY}, 
thus leaving the LF vacuum trivial.
Now that the vacuum is trivial, whole information about 
the SSB in the LF quantization should reside in the operator 
instead of the state, namely, in 
the zero mode whose dynamics is governed by the zero-mode constraint. 

However, direct application of the zero-mode constraint leads to 
inconsistent result: The DLCQ allows neither the 
emission vertex 
of the NG boson 
nor the corresponding current 
vertex, as far as the NG boson mass is exactly zero, 
or conservation of the LF charge 
is imposed (``(false) no-go theorem" ) \cite{KTY}. 
Namely,
{\it solving the zero-mode constraint does not give rise to SSB at all
in
the exact symmetric case} $m_{\pi}^2\equiv 0$, in contradiction to
the naive expectation mentioned above \cite{HKSW,Rob,BPV}.

In order to recover the NG phase in DLCQ with the trivial vacuum, 
we thus need to formulate 
 non-conservation of
the LF charge.
We propose \cite{KTY} that it can be achieved in DLCQ by first 
introducing explicit-symmetry-breaking mass of the 
NG boson $m_{\pi}$ and then taking 
its massless limit. This mass plays a role of regularization of the 
infrared singularity of the zero mode. 
Based on the notion of PCAC, it will be clarified how the SSB 
without NG-boson mass becomes self-contradictory in
DLCQ and how the arguments leading to the above ``(false) no-go theorem" 
went wrong. 
We find that the NG phase on the LF is characterized   
by the singular behavior of the NG-boson zero mode: The global 
zero mode (zero mode integrated over the LF) of the NG boson 
 must be proportional to $1/m_{\pi}^2$ in the symmetric limit 
 $m_{\pi}\rightarrow 0$ \cite{KTY}. 
This in fact leads to 
{\it non-conservation of the LF charge while preserving
the trivial vacuum.}

The above general feature of the SSB on the LF will be further
demonstrated in
a concrete field theoretical model, 
the linear sigma model, in which the role of the zero mode 
is most explicitly illustrated.  We derive coupled zero-mode 
constraints with the NG-boson mass included and solve them 
in perturbation around the classical broken solution
which corresponds to the classical broken vacuum in the 
equal-time quantization.
 The singular behavior of the NG-boson zero mode 
is indeed explicitly demonstrated by such a perturbative solution
which at tree level yields non-vanishing $\sigma\pi\pi$ and NN$\pi$ 
vertices consistently with the usual result of the equal-time
quantization.
It is most remarkable that 
the current conservation or the LF-charge conservation 
actually breaks down due to such a singular behavior of the NG boson
zero
mode. 

We also note that were it not for the NG boson mass from the onset, 
the zero-mode constraints, after integration over the 
transverse coordinate, would take an essentially 
the same form as that of 
the two-dimensional massless scalar theory which, however,
will be shown to be ill-defined in accord with 
Coleman's theorem \cite{Cm}.  
Thus the LF theory without NG-boson mass in 
four dimensions is also ill-defined and hence introduction 
of the NG-boson mass  
and the resulting non-conservation of the LF charge 
is inevitable in DLCQ. 

Finally, we shall compare our result with the 
zero mode problem in the continuum theory. 
In the continuum theory it is rather difficult to 
remove the zero mode
 in a sensible manner as was pointed out by
Nakanishi and Yamawaki \cite{NY} long time ago: 
The real problem is {\it not a single mode} with 
$p^+\equiv 0$ (which is merely of zero measure and harmless)
but actually the {\it accumulating point} $p^+\rightarrow 0$ 
as can been seen 
from $1/p^+$ singularity in the Fourier transform of the 
sign function $\epsilon (x^-)$ appearing in
the canonical commutator on LF. This prevents us 
from constructing even a free theory on the LF 
({\it no-go theorem}\cite{NY}), 
which actually 
cannot be  overcome either by taking the simple 
continuum limit $L\rightarrow \infty$ of DLCQ nor any other 
existing method \cite{NY}.
Besides this most difficult problem,
in this paper we shall point out another problem 
in the continuum theory. Namely,
the {\it SSB charge on LF does not annihilate the vacuum}, 
if we formulate the sigma model with 
SSB on the LF with careful treatment of the boundary condition.
Even if we pretend to have removed the zero mode, 
{\it as far as the canonical commutator 
takes the form of the sign function} $\epsilon (x^-)$, it
inevitably leads to a nontrivial vacuum, namely, 
the LF charge does not annihilate
the vacuum. This in fact corresponds to difficulty to remove 
the zero mode as the accumulating point mentioned above (in
contradiction 
to a widely spread expectation \cite{Wils}).  

We then
suggest that a possible way out of this problem within
the continuum theory would be the ``$\nu$-theory'' proposed by  
Nakanishi and Yamawaki \cite{NY} which removes 
the zero mode in the continuum theory by shaving the vicinity of 
the zero mode in such a way that 
{\it the sign function}
in the commutator is
{\it modified}
to a certain function 
vanishing at $x^- =\pm \infty$. This theory is 
expected to yield the same 
result as that we obtain in this paper based on DLCQ, although
it does not
overcome the no-go theorem \cite{NY} 
mentioned above. 
The no-go theorem and the $\nu$-theory
will be further discussed in great detail in order
to remind the reader of the old results \cite{NY}. 

The plan of this paper is as follows:
In Sec.2 we recapitulate 
the canonical formalism of DLCQ for the  
scalar theory and the zero-mode constraint 
in a way slightly different 
from the original one \cite{MY}. 
In Sec.3 we consider the SSB of the continuous symmetry and 
show how the NG phase can be realized in the trivial 
LF vacuum through the explicit-symmetry-breaking mass of the 
NG boson. The singular behavior of the global zero mode 
of the NG boson in the $m_{\pi}\rightarrow 0$ is required, 
which implies non-conservation of the SSB current.      
In Sec.4 we apply our formulation 
to the linear sigma model by treating the 
zero-mode constraints explicitly.  
It is shown that the tree-level amplitude of both the 
$\sigma\pi\pi$ and the  
$NN\pi$ scatterings  are actually obtained in DLCQ due to  
the singular behavior 
of the NG-boson zero mode in the symmetric limit. 
In Sec.5 we discuss the zero-mode problem
in the continuum theory, which is quite different from
that in DLCQ. 
Sec.6 is devoted to the Summary 
and Discussions where we present particular discussions
on the no-go theorem  
and the $\nu$-theory in detail.  
A detailed derivation of the DLCQ canonical commutator is given in
Appendix A.
In Appendix B the special status of the boundary condition on the LF 
is discussed. In Appendix C, we describe the unbroken phase 
of $O(2)$-symmetric linear sigma model in DLCQ 
and check the operator ordering (Weyl ordering)
we use in discussing the perturbative solution. 
The property of the higher order perturbative 
solutions  of the zero-mode constraints is studied 
in Appendix D.


\section{Zero Mode in the Discretized Light-Cone Quantization} \par
\my
In this section  we review for later purpose 
the canonical DLCQ of the scalar theory 
\cite{MY}. 
Throughout this paper 
we use the convention of the LF coordinate 
$x^{\mu}=(x^+,\vec{x})=(x^+,x^{\bot},x^-)$,
where 
\begin{eqnarray}
x^{\pm}&\equiv&\frac{1}{\sqrt 2}(x^0 \pm x^3) \quad,\\
x^{\bot} &\equiv& (x^1,x^2) \quad.
\end{eqnarray}
The quantization 
surface on the equal ``LF time" $x^+$ is defined
in the finite region 
$-L\leq x^- \leq L$ \cite{MY,casher,PB}, 
while no such restriction is necessarily imposed for the 
transverse coordinates $x^{\bot}$. 
The ``continuum" limit $L \rightarrow \infty$ 
(or, more precisely, infinite volume limit) is 
taken at the final 
stage of the whole calculations. 
 We use the notation 
$\int d^3\vec{x}\equiv \lim_{L \rightarrow \infty}
\int_{-L}^{L}dx^-d^2x^{\bot}$.

Let us consider the self-interacting scalar 
theory in four dimensions 
whose Lagrangian is 
expressed in terms of the LF coordinate as  
\begin{equation}
{\cal L}={\partial}_{+} \phi {\partial}_{-} \phi 
-\frac{1}{2}({\partial}_{\bot}\phi)^2-\frac{1}{2}
{\mu}^2 {\phi}^2-V(\phi)
\quad , 
\label{sl}
\end{equation}
where $V(\phi)$ is a potential. 
The canonical momentum conjugate to $\phi(x)$ is 
\begin{equation}
\pi(x)=\frac{\partial {\cal L}}{\partial({\partial}_+ \phi)}
=\partial_{-}\phi(x) 
\quad ,  
\end{equation}
which leads to a primary constraint of the theory:
\begin{equation}
\Phi(x)=\pi(x)-\partial_{-}\phi(x)\approx 0
\quad .  
\label{primary}
\end{equation}

Since $x^-$ is restricted to the finite region, 
the boundary condition should be specified at 
$x^{-}=\pm L$. We adopt the periodic boundary condition 
on $x^-$ \cite{MY}, which is consistent with non-vanishing 
vacuum expectation value of the scalar field. 
In fact, very existence of the zero mode is related to 
this periodic boundary condition. Other boundary conditions 
such as the anti-periodic one 
will be discussed in Appendix B. 
Owing to the boundary condition in the finite box, 
all surface terms can be treated unambiguously, 
while their treatment is subtle in the continuum framework 
(see Sect. 5 and Appendix B).  
Actually, as was emphasized by Steinhardt \cite{STH},  
the boundary condition should always be specified even in the
``continuum '' theory
in order to have a consistent LF quantization.
In fact, the boundary condition on LF includes 
a part of the dynamics in sharp contrast to 
the equal-time quantization. That is, different boundary condition
defines a different 
theory. We shall clarify a special role of the 
boundary condition on the LF in Appendix B. 

Since the zero mode in DLCQ is clearly separated from other modes,  
we may make an orthogonal decomposition of 
the primary constraint into two parts as follows \cite{H}.
Let us divide the scalar field $\phi(x)$ into  
the oscillating modes $\varphi (x)$ plus the zero mode $\phi_0
(x^+,x^{\bot})$:
\begin{equation}
\phi(x)=\varphi(x)+\phi_0(x^+,x^{\bot})
\quad , 
\label{dec-1} 
\end{equation}
\begin{equation}
\phi_0\equiv \frac 1{2L}\int_{-L}^{L}\phi(x)dx^{-} \quad .
\end{equation}
The conjugate momentum 
$\pi$ may also be divided as 
\begin{equation}
\pi(x)=\pi_{\varphi}(x)+\pi_0 (x^+,x^{\bot})
\quad ,  
\label{dec-2}
\end{equation}
where $\pi_0$ and $\pi_{\varphi}$ are the zero modes conjugate to 
$\phi_0$ and that to the remaining orthogonal part $\varphi(x)$,
respectively. 
Now, substituting (\ref{dec-1}) and (\ref{dec-2}) into 
(\ref{primary}), we have two independent 
constraints, 
\begin{equation}
\Phi_{1}(x)\equiv \pi_{\varphi}(x)-\partial_{-}
\varphi(x)\approx 0
\quad
\label{Phi1} 
\end{equation}
and 
\begin{equation}
\Phi_{2}(x)
\equiv \pi_{0}(x^+,x^{\bot})\approx 0
\quad ,  
\label{Phi2}
\end{equation}
in place of the original one (\ref{primary}).

 From the fundamental Poisson bracket 
\begin{equation}
\{ \phi(x),\pi(y)\}
=\delta^{(3)}(\vec x-\vec y) \quad, 
\label{poisson1}
\end{equation}
we obtain   
\begin{equation}
\{\phi_0,\pi_0\}=\frac 1{2L}\delta^{(2)}(x^{\bot}-y^{\bot}), 
\quad \{\phi_0,\phi_0\}=\{\pi_0,
\pi_0\}=0
\quad , 
\label{poisson2}
\end{equation}
and 
\begin{equation}
\{\varphi(x),\pi_{\varphi}(y)\}=\left\{\delta(x^-
-y^-)-\frac1{2L}\right\}
\delta^{(2)}(x^{\bot}-y^{\bot})\quad,
\label{poisson3}
\end{equation}
\begin{equation}
\{\varphi(x),\varphi(y)\}=\{\pi_{\varphi}(x),\pi_{\varphi}(y)\}=0 
\quad,
\label{poisson4}
\end{equation}
where $x^+=y^+$ is understood.
All other Poisson brackets are equal to zero as expected.   

The total Hamiltonian is obtained by adding the primary constraints
to the canonical one $H_c$: 
\begin{eqnarray}
H_T &\equiv& H_c+\int d^3 \vec{x}\hspace{5pt} 
\left[ v_1(x)\Phi_1(x)+
v_2(x) \Phi_2(x) \right] \quad,\\
H_c&=& \int d^3 \vec{x} \hspace{5pt}
\left[\frac{1}{2}\{(\partial_{\bot}\phi)^2 
+\mu^2 \phi^2\}+V(\phi)\right]
\quad , 
\end{eqnarray}
where $v_2$ and $v_1$ are the zero mode and 
the remaining part of the Lagrange multiplier, respectively. 
The multiplier $v_1$ is determined by the consistency condition 
for $\Phi_1(x)$ through the 
relation
\begin{equation}
\partial_{-}v_1(x)=\frac 1{2}\{ \pi(x),H_c \}-\frac 1{2}\{ \pi_0,H_c\}
\quad ,  
\end{equation}
which can be easily integrated without ambiguity owing to the periodic
boundary 
condition. 
On the other hand, the consistency condition for 
$\Phi_2(x)$ 
\begin{equation}
\dot {\Phi}_2(x)=\{\Phi_2(x),H_T\}
={1 \over 2L}\int_{-L}^L dx^{-}
\left[(\partial_{\bot}^2-\mu^2)\phi-
{\partial V \over \partial \phi}\right] \approx 0
\quad 
\end{equation}
leads to a new constraint so-called 
``zero-mode constraint"  \cite{MY}: 
\begin{equation}
\Phi_3 (x) \equiv \frac{1}{2L}\int_{-L}^L dx^{-}
\left[(\mu^2-\partial_{\bot}^2)\phi+\frac{\partial V}{\partial
\phi}\right]  
\quad . 
\label{zmconstraint}
\end{equation}
The consistency condition for the zero-mode constraint 
yields no further constraint and just 
determines the multiplier $v_2$. 
Note that in deriving these relations we have used the condition
\begin{equation}
\delta(x^--L)=\delta(x^-+L)
\quad , 
\end{equation}
which comes from the definition of the delta function with 
the periodic boundary condition:
\begin{equation}
\delta(x^-)=\frac{1}{2L}\sum_{n \in {\bf Z}} e^{\frac{in\pi}{L}x^-}
\quad .  
\end{equation}

Having obtained all the second-class 
constraints, we are ready to calculate the  
Dirac bracket of two arbitrary 
dynamical variables $A(x)$ and $B(y)$ as 
\begin{eqnarray}
\{A(x),B(y)\}_{DB} &\equiv& \{A(x),B(y)\} \nonumber \\ 
                   &-&\sum_{i,j}\int d^3 \vec{u} \int d^3 \vec{v} 
                   \{A(x),\Phi_i(u)\}
                     (C^{-1})_{i,j}(u,v)\{\Phi_j(v),B(y)\} 
\quad ,  
\label{DM}
\end{eqnarray}
where $ (C^{-1})_{i,j} $ is the inverse of 
$C_{i,j}(x,y)\equiv \{\Phi_i(x),\Phi_j(y)\}$ which is
the matrix of Poisson bracket of the constraints.
The inverse matrix can be calculated by noting the separation of 
the zero mode from other modes. For instance, the matrix element 
\begin{eqnarray}
C_{1,1}(x, y)\equiv \{\Phi_1 (x),\Phi_1(y)\}&=&(\partial^{y}_{-}
-\partial^{x}_{-})\delta(x^- -y^-)\cdot\delta^{(2)}(x^{\bot}-y^{\bot})
\nonumber \\
&=&\frac{1}{2L}\sum_{n\in {\bf Z}}\left(\frac{-2in \pi}{L}\right)
\hspace{5pt}e^{\frac{in\pi}{L}
(x^- -y^-)}
\cdot\delta^{(2)}(x^{\bot}-y^{\bot})
\quad , 
\end{eqnarray}
has its inverse 
\begin{eqnarray}
[C_{1,1}(x, y)]^{-1}
&=&-\frac{1}{4}\left\{\epsilon(x^--y^-)-\frac{x^--y^-}{L}\right\}
\cdot\delta^{(2)}(x^{\bot}-y^{\bot})
\nonumber 
\\
&=&\frac{1}{2L}\sum_{n \ne 0}\left(\frac{-L}{2in\pi}\right)\hspace{5pt}
e^{\frac{in\pi}{L}(x^--y^-)}
\cdot\delta^{(2)}(x^{\bot}-y^{\bot})
\quad ,  
\label{inverse}
\end{eqnarray}
in the sense that 
\begin{equation}
\int^L_{-L}dz^{-} \int dz^{\bot} C_{1,1} (x,z) [C_{1,1}(z,y)]^{-1}=
                 \frac{1}{2L}\sum_{n \ne 0, n \in {\bf
Z}}e^{i\frac{n\pi} 
                {L}(x^{-}-y^-)}
\cdot\delta^{(2)}(x^{\bot}-y^{\bot})
\quad , 
\label{inverse2}
\end{equation}
where $\epsilon(x)$ is the sign function satisfying  
$\partial_x \epsilon(x)=2\delta(x)$. 
Note that the right hand side (r.h.s.)  
in (\ref{inverse2}) is a delta function minus zero-mode contribution
as it should, since the zero mode is already
subtracted from $\Phi_1$ beforehand.

After the Dirac bracket is taken, all the second-class constraints
become
strong relations and so is the zero-mode constraint
(\ref{zmconstraint}):
\begin{equation}
\frac{1}{2L}\int_{-L}^L dx^{-}
\left[(\mu^2-\partial_{\bot}^2)\phi+\frac{\partial V}{\partial
\phi}\right]  =0
\quad ,
\label{zmconstraint2}
\end{equation}
which is further converted into the operator relation, when we pass to
the quantum
theory via the correspondence principle between the Dirac bracket and
the commutator,
$\{\quad \}_{DB}\rightarrow -i[\quad]$. 
This implies that the zero mode  
is {\it not an independent degree of freedom} but is
implicitly written in terms of other oscillating modes.
It was actually the central issue of Maskawa and Yamawaki \cite{MY}
who claimed that 
{\it such a constrained zero mode can in principle be
solved away out of the physical Fock space 
and hence the trivial LF vacuum is justified in DLCQ}
. 
It is also noted \cite{MR} that the zero-mode constraint
(\ref{zmconstraint2})
can also be obtained by simply
integrating in $x^-$ the Euler-Lagrange equation 
$(\Box +\mu^2)\phi= -\frac{\partial V}{\partial \phi}$ 
with use of the periodic boundary condition:
\begin{equation}
0=-\int^{L}_{-L}dx^{-}2\partial_{+}\partial_{-}\phi
 =\int^{L}_{-L}dx^{-}\left[(\mu^2 -\partial_{\bot}^2)\phi +
 \frac{\partial V}{\partial \phi}\right] \quad.
\label{zmconstraint3}
\end{equation} 
Namely, the zero mode constraint is a part of the 
equation of motion and 
the zero mode is nothing but an auxiliary 
field having no kinetic term.

  Through the above correspondence principle, 
we obtain from (\ref{DM}) 
 the canonical commutation relation \cite{MY}:
\begin{equation}
\left[\varphi(x),\varphi(y)\right]=-{i \over
4}\left\{\epsilon(x^--y^-)-{x^--y^- 
\over L}\right\}\delta^{(2)}(x^{\bot}-y^{\bot})\quad 
\label{comm}
\end{equation}
for the field {\it without zero mode},
which is a direct consequence of (\ref{inverse}).
In sharp contrast to the sign function 
in the continuum theory (see Sect.5),
here in DLCQ we observe 
presence of the extra term $(x^- - y^-)/L$ in the commutator
(\ref{comm}), 
which is nothing but a term subtracting the zero mode 
as can be seen from (\ref{inverse}).
Note that (\ref{comm}) is the same as the commutation relation of 
the full field $\phi$ in the free theory in which  
the zero mode becomes 
identically zero through the zero-mode constraint (\ref{zmconstraint2}).
 
By computing (\ref{DM}) for the full matrix, we further obtain a
commutation 
relation for the {\it full field} including the zero mode 
(see Appendix A): 
\begin{eqnarray}
\left[\phi(x),\phi(y)\right] &=& -\frac{i}{4}\int du^{-} dv^{-}
\left[\beta(x^{\bot}) \alpha(u^-,x^{\bot})
                    -\delta(u^{-}-x^{-})\right] 
                      \left\{\epsilon(u^--v^-)-
                      \frac{u^--v^-}{L} \right\}\times \nonumber\\
                  & & 
\left[\beta(y^{\bot})\alpha(v^-,y^{\bot})-\delta(v^{-}-y^{-})\right]
\delta^{(2)}
                     (x^{\bot}-y^{\bot}) \quad,
\label{comm1}
\end{eqnarray}
where 
\begin{equation}
\alpha(\vec{x})
\equiv\mu^2-\partial_{\bot}^2+\frac{\partial^2 V}{\partial \phi^2}
\quad, 
\quad \quad 
\beta^{-1}(x^{\bot})
\equiv\int^L_{-L} dx^{-}
\alpha(\vec{x})
\quad . 
\label{alphabeta}
\end{equation}
At first sight, (\ref{comm1}) looks different 
from the original expression in \cite{MY}:
\begin{equation}
\left[\phi(x),\phi(y)\right]
=-\frac{i}{4}\left\{\epsilon(x^--y^-)-2\beta \int_{y^-}^{x^-}
                   \alpha(z^-)dz^- \right\}\delta^{(2)}
                     (x^{\bot}-y^{\bot})
\quad , 
\label{comm2}
\end{equation}
which was obtained 
without orthogonal decomposition of the primary constraint 
(\ref{primary}) into two parts,
zero mode and non-zero modes. 
However, explicit computation shows
(Appendix A)
that both are  in fact equivalent to each other 
if the operator ordering is disregarded (i.e., in the sense of Dirac
brackets). 
Note that $\alpha$ and $\beta$ contain field operators in the
interacting 
theory and hence (\ref{comm1}) (or (\ref{comm2})) is generally a 
complicated {\it operator-valued commutation relation},
which is nothing but a consequence 
of the constrained zero mode carrying the vital dynamical information 
of the theory.  

Here we should remark on the operator ordering to be consistent 
with the Dirac quantization. In the Dirac procedure, 
the Dirac bracket is constructed so that
all the second class constraints can automatically hold as strong 
identities. 
This property must be preserved in passing from the 
classical theory to the quantum one. 
In the case at hand, the zero-mode constraint should commute with 
any operator just by calculation using 
the commutator for the full scalar fields.  
Rather such operator 
orderings in the quantum theory must be determined for both the
zero-mode constraint and 
the r.h.s. of (\ref{comm1}) (or (\ref{comm2})) simultaneously. 
However this is an extremely difficult task and we take a different
approach:
Instead of a requirement for the zero-mode constraint to be 
a strong operator identity, we assume the Weyl ordering for the
operators in
the zero-mode constraint to solve it explicitly.
The solution of the 
zero mode then leads to the commutator (\ref{comm1})  
with a definite operator ordering. 
Our choice of the Weyl ordering is based on 
the general argument \cite{weyl}. 
Moreover it will be 
justified through the study of linear sigma model 
in the section 4 and Appendix C. 

\section{Nambu-Goldstone Boson on the Light Front}\par
\my

It is now widely believed that non-perturbative phenomena 
due to nontrivial vacuum  
in equal-time quantization can be understood in the LF quantization 
through the operator property of the zero mode. 
Among the various zero modes, we focus on the bosonic 
zero mode which has been shown to be a dependent degree of freedom 
and is expected to play a key role to realize SSB on the LF. 
One might then  expect
that the non-perturbative vacuum structure in equal-time 
quantization is simply replaced by the solution of the 
zero-mode constraint.  However the problem 
is not so simple for the realistic case with continuous symmetry in 
four dimensions, 
whose realization is usually associated with the massless NG boson.

The purpose of this 
section is to propose   
the criteria for global continuous symmetry breaking on 
the trivial LF vacuum \cite{KTY}. 
Before drawing our main conclusion, we first show that 
the naive application 
of the zero-mode constraint will not lead to the NG phase 
at all in contradiction to the above expectation
 ("(false) no-go theorem" \cite{KTY}):
{\it If the zero-mode constraint is not reguralized by 
the explicit mass of the NG boson, 
the NG phase cannot be realized 
in DLCQ. Namely, there is no coupling of NG boson as well as no 
associated current vertex at $q^2=0$, where $q^{\mu}$ 
is the momentum of the NG boson. }

In order to confirm our  assertion, let us start with 
assuming that the NG phase is already 
realized on the LF in the presence of an  exactly 
conserved current and examine its consequence. Consider  
the arbitrary NG-boson emission process  $A \rightarrow B + \pi$, 
where both A and B represent one-particle states which couple with 
the NG boson.  The 
index of the NG boson associated with 
the
internal 
symmetry is omitted for
 simplicity.    

Based on the reduction formula, the transition amplitude may be 
written as 
\begin{eqnarray}
\lefteqn{\langle B, \pi(q)\vert A \rangle \equiv i 
\int d^4 x e^{iqx} \langle B 
\vert \Box \pi(x) \vert A \rangle} \nonumber \\
&=&i(2\pi)^4 \delta(p_A^{-}-p_B^{-}-q^-)
\delta^{(3)}(\vec {p}_A-
\vec{p}_B-\vec{q})
\langle B \vert j_{\pi}(0)\vert A 
\rangle \quad ,   
\end{eqnarray}
where $\pi(x)$ 
and $j_{\pi}(x)=\Box \pi(x)=(2\partial_{+}\partial_{-}
-\partial_{\bot}^2) \pi(x)$
are the interpolating field of NG boson, which is exactly 
massless, 
and
the source function of the NG boson, respectively, and
$q^{\mu}=p^{\mu}_A-p^{\mu}_B$ 
is the NG-boson four momenta. 

Taking the collinear momentum frame \cite{Weinberg}, $q^+=q^{\bot}=0$
and $q^{-}\ne 0$  {\it which is not soft momentum}  
for the on-shell  
NG boson with $q^2=0$, we find 
that the NG-boson emission vertex does
vanish as follows:
\begin{eqnarray}\label{emission}
\lefteqn{(2\pi)^3\delta^{(3)}(\vec{p}_{A}-\vec{p}_{B})
\langle B \vert j_{\pi}(0)\vert
A \rangle}\nonumber\\
&=&\int d^{3}\vec{x} \langle B
\vert\left( 2\partial_{+}\partial_{-}-\partial_{\bot}^2\right)\pi 
\vert A
\rangle\quad\nonumber
\label{re}\\
&=&\int d^{2}x^{\bot}\lim_{L\rightarrow\infty}\langle B \vert 
\Bigl(\int^{L}_{-L}dx^{-}2\partial_{+}\partial_{-}\pi\Bigr)
\vert A
\rangle=0 \quad ,
\end{eqnarray}
where the periodic boundary condition was used 
for the NG boson field $\pi$ as before.  
As seen from (\ref{zmconstraint3}),  
the last line is nothing but a zero-mode 
constraint for the massless field, and hence {\it the zero-mode
constraint itself 
dictates that the NG boson vertex should vanish}. Thus we have
established that 
the solution of the zero-mode 
constraint, 
whether perturbative or non-perturbative 
or even exact, does not lead  
to the NG phase at all. 

Another symptom of this disease 
is the vanishing of the current vertex 
for the SSB current as a direct consequence of the LF charge
conservation,
which
again comes from our periodic boundary condition through the local
current conservation. 
The current vertex  
is an analogue of $g_A$ for the nucleon matrix 
element and is related 
to the NG boson vertex ($g_{NN\pi}$ for the nucleon case) in the usual 
SSB argument 
through the analogue of the Goldberger-Treiman relation. 
(Caveat for the nucleon case will be given later.)  Now that we have
seen that
 the NG boson vertex vanishes
due to the periodic boundary condition,
we may naturally guess that the current vertex also 
should for the same reason.
In what follows we  shall argue that this is indeed the case.
In the NG phase the  
current $J_{\mu}$ is divided into the pole term consisting of  
an interpolating field of the NG boson and the remaining  
non-pole term, that is, 
\begin{equation}
J_{\mu}=-f_{\pi}\partial_{\mu}\pi+\widehat
J_{\mu}\quad,
\end{equation}
where $f_{\pi}$ is the ``decay constant'' of 
the NG boson 
and $\widehat J_{\mu}$ denotes the non-pole term which yields the
current 
vertex.
Now, integrating the local current conservation over the LF, 
we find that the NG-boson pole term drops out, 
leaving only the non-pole term due to the 
periodic boundary condition as before. 
Then we establish the vanishing current vertex
as follows \cite{KTY}:  
\begin{eqnarray} 
\lefteqn{0=\langle B \vert 
\int d^3\vec{x}\, \partial_{\mu}\widehat J^{\mu}(x)\vert A
\rangle_{x^+=0}}\nonumber\\
&=&-i(2\pi)^3\delta^{(3)}(\vec{q})\, \displaystyle{
\frac{m_{A}^2-m_{B}^2}{2p_A^+}}\langle B \vert \widehat J^+(0)
\vert A \rangle \quad ,
\label{c-vertex} 
\end{eqnarray} 
where $q^-=p_A^--p_B^-=(m_A^2-m_B^2)/2p^{+}_A$, which implies 
\begin{equation}
\langle B\vert\widehat J^{+}(0)\vert A\rangle=0\quad,
\end{equation}
as far as $m^2_{A}\ne m^2_{B}$.\footnote{
For the case where the two particles A and B have 
a degenerate mass, i.e. 
$m_{A}=m_{B}$, Eq.(\ref{c-vertex}) by itself 
does not implies the vanishing current vertex
$\langle B\vert\widehat J^{+}(0)\vert A\rangle=0$. However, 
in this case $q^2\rightarrow0$ 
corresponds to the soft momentum limit $q^{\mu}\rightarrow 0$, 
which implies that even in the usual equal-time treatment, 
the NG-boson emission vertex vanishes 
anyway by the low energy theorem, even when the current vertex
is non-zero (for the nucleon case, 
the current vertex is also zero for kinematical
reason, although $g_A$ is non-zero, see the discussion in Sect.4)
\cite{BKY}.
}
The current vertex $\langle B\vert\widehat J^{+}(0)\vert A\rangle$ 
at $q^2=0$ is nothing but the matrix element of LF charge 
$\widehat Q\equiv \int d^3\vec{x}\, \widehat J^{+}$ 
constructed only from the non-pole term
(well-defined charge even in the SSB phase)
and is essentially the same as ``$X$ matrix'' 
of Weinberg \cite{Weinberg}. The chiral algebra of LF charge 
actually   yields the celebrated  Adler-Weisberger 
sum rule \cite{AW} and its extensions as an algebraic realization 
of the chiral symmetry in terms of the notion of representation 
mixings among hadronic states with non-degenerate masses $m_A^2\ne
m_B^2$
\cite{Weinberg,GH,Ida}.
Hence the vanishing of the current vertex 
invalidates whole success of the
Adler-Weisberger sum rules and the 
associated representation mixings. Actually, 
vanishing of the 
current vertex means 
conservation of the LF charge
$\widehat Q$ which immediately follows from conservation of 
the full LF charge $Q\equiv \int d^3 \vec{x}\ J^{+}$, 
since $Q$ always reduces to $\widehat Q$, 
with the pole part being dropped 
out of $Q$ due to the integration over the LF:    
\begin{equation}
[\widehat Q, P^-]=[Q, P^-]=0
\quad.
\end{equation}
We again emphasize that conservation of the LF charges
is a direct consequence of the periodic
boundary condition we are using,
provided that the local current is conserved as 
we imposed.

So, what went wrong?  
One might use other boundary conditions than the periodic one.
In Appendix B we shall argue that beside the periodic boundary 
condition, only the anti-periodic one can be consistent in DLCQ, 
which however yields no SSB because
of obvious absence of the zero mode.  
One might then give up DLCQ and consider the continuum
theory from the onset, in which case, however, 
we still need to specify the boundary condition 
in order to have a consistent LF theory
\cite{STH} as
we shall discuss also in Appendix B. 
The best we can do in the continuum theory 
will be described in Sect. 5, which, although 
can give non-zero NG boson 
vertex and current vertex (i.e., non-conservation of the LF charge)
due to the boundary condition, will result in another disaster, 
namely, the LF charge does not annihilate the vacuum, 
thus invalidating the trivial vacuum as the
greatest advantage of the whole LF
approach. One also might suspect that the finite volume in $x^-$
direction in DLCQ
could be the cause of this NG-boson decoupling, 
since it is well known that SSB does not occur in the 
finite volume. However, 
we actually take the $L \rightarrow \infty$ limit in the end, 
and such a limit 
in fact must realize SSB as was demonstrated in the equal-time 
quantization in the infinite volume limit of the finite box 
quantization \cite{NJL}. Moreover, 
in the case at hand in four dimensions, 
the transverse directions $x^{\bot}$ 
extend to infinity.
Hence 
this argument is totally irrelevant 
any way. 

Therefore the above result is {\it not 
an artifact} of the periodic boundary condition 
and DLCQ but is deeply 
connected to the very nature of the LF quantization, 
namely the zero mode.
Thus, as far as the trivial property 
of the LF vacuum is to be maintained, 
the only way to recover the NG phase  seems to break the 
symmetry explicitly. 
By the various arguments to follow along this line 
both in this section and
Sect.4, we actually conclude \cite{KTY}: 

{\it  The NG phase can be realized in DLCQ only when  the 
NG-boson mass $m_{\pi}$ is introduced into  the theory. 
The non-vanishing NG-boson emission vertex as well as 
the corresponding 
current vertex at $q^2=0$ is recovered through 
the 
explicit
symmetry-breaking 
term in the  
$m_\pi \rightarrow 0$ limit,  which is characterized by the 
singular behavior of   
the NG-boson zero mode $\omega_{\pi}$: }
\begin{equation}
\int d^3 \vec{x}\, \omega_{\pi}\sim \frac 1{m_{\pi}^2} 
\quad .
\label{omega} 
\end{equation}
\vspace{5mm} 

We can easily confirm (\ref{omega})   
with the help of the PCAC hypothesis: 
$\partial_{\mu}J^{\mu}(x)=f_{\pi}m_{\pi}^2\pi(x)$.   
Since the current divergence of the non-pole term 
$\widehat J^{\mu}(x)$ reads $\partial_{\mu}\widehat J^{\mu}(x)
=f_{\pi}(\Box+m_{\pi}^2)\pi(x)= 
f_{\pi}j_{\pi}(x)$, we obtain 
\begin{eqnarray}
\langle B \vert \int d^3\vec{x}\,
\partial_{\mu}\widehat J^{\mu}(x)
\vert A \rangle 
&=&f_{\pi}m_{\pi}^2
\langle B \vert 
\int d^3 \vec{x}\, \pi(x)\vert A \rangle \label{PCAC1}\quad  \\  
&=&
\langle B \vert 
\int d^3 \vec{x}\, f_{\pi} j_{\pi}(x) 
\vert A \rangle \quad , \label{PCAC}
\end{eqnarray}
where the integration of the pole term $\Box \pi(x)$ is dropped out 
as before. On the r.h.s. of (\ref{PCAC1}), one can replace 
$\pi(x)$ by the zero mode 
$\omega_{\pi} \equiv \frac{1}{2L}\int_{-L}^{L} dx^- \pi(x)$  
because the oscillating modes drop out due to the periodic 
boundary condition. 
Suppose that $\int d^3\vec{x}\, \omega_{\pi} (x)
=\int d^3\vec{x}\, \pi (x)$ is regular when  
$m_{\pi}^2\rightarrow 0$, this does not lead to the NG phase 
at all, because the remaining two terms 
then
become vanishing. 
In order to have the non-zero NG-boson emission 
vertex (\ref{PCAC}) 
as well as the non-zero current vertex (l.h.s. of (\ref{PCAC1})) 
at $q^2=0$, the zero mode $\omega_{\pi}$ must 
behave as (\ref{omega}). 

This implies that at the quantum level 
the LF charge $Q=\widehat Q$ is not conserved 
or the current conservation does not hold
for its particular Fourier component with $\vec{q}=0$ even 
in the symmetric limit: 
\begin{equation}
\dot{Q}=\frac{1}{i}[Q, P^{-}]=\partial^{\mu}
J_{\mu}\vert_{\vec{q}=0}=f_{\pi} 
\lim_{m_{\pi}^2\rightarrow 0} 
m^2_{\pi}\int d^3 \vec{x}\, \omega_{\pi} \neq 0 \quad ,
\label{nonconserv}
\end{equation}
although  we can recover the conserved current 
at the classical level. 

The situation may well be clarified when we consider the 
general expression  
for the current matrix element in momentum space
with an explicit symmetry breaking: 
\begin{equation}
\displaystyle{
\frac{m_{\pi}^2f_{\pi}j_{\pi}(q^2)}{m_{\pi}^2-q^2}=
\partial^{\mu} J_{\mu}(q)=
\frac{q^2f_{\pi}j_{\pi}(q^2)}{m_{\pi}^2-q^2}
+\partial^{\mu}\widehat J_{\mu}(q) \quad ,
}
\label{PCAC-mom}
\end{equation}
 which is a weaker condition than the operator 
relation of PCAC hypothesis. What we have done 
in proving the absence of NG phase for the exactly conserved current 
(``(false) no-go theorem'')
is summarized as follows: 
We first set the l.h.s. of (\ref{PCAC-mom}) to
zero or equivalently, assumed implicitly the regular 
behavior of $\int d^3\vec{x}\, \omega_{\pi}(x)$ in the massless limit 
in accord with the current conservation 
$\partial^{\mu} J_{\mu}=0$. Second, 
the first term (NG-boson pole term) on the r.h.s. of (\ref{PCAC-mom})
vanishes 
rigorously  due 
to the periodic boundary 
condition or the zero-mode constraint in the DLCQ with $\vec{q}=0 $
$(q^2=0)$. 
Thus we arrived at  $\partial^{\mu}\widehat J_{\mu}(q)=0$ in addition 
to the vanishing of the NG-boson vertex.
However, this procedure is
equivalent to claiming
the nonsense relation 
\begin{equation}
1=\lim_{m_{\pi}^2,\,q^{2}\rightarrow 0}
(\frac{m^{2}_{\pi}-q^{2}}{
m^{2}_{\pi}-q^{2}})=0 \quad, 
\end{equation}
as far as $f_{\pi}j_{\pi}\ne 0$ (NG phase).
Therefore the ``$m_{\pi}^2 \equiv 
0$'' theory with vanishing l.h.s. 
is ill-defined in DLCQ and 
we should define the symmetric  
limit after introducing the 
explicit
symmetry-breaking 
term.


\section{The Sigma Model}\par 
\my

Based on DLCQ discussed in Sect.2, 
let us now  demonstrate (\ref{omega}) 
by explicitly solving the zero-mode constraints with the 
NG-boson mass in a concrete model theory \cite {KTY}. 
As the simplest but a nontrivial example,   
we consider $O(2)$-symmetric linear sigma model defined by the 
Lagrangian: 
\begin{eqnarray}
{\cal L}&=&\bar{\psi}i\gamma^{\mu}\partial_{\mu}\psi-g\bar{\psi}
(\sigma+i\gamma_{5}\pi)\psi 
+ \frac{1}{2}(\partial_{\mu}\sigma)^2+\frac{1}{2}
(\partial_{\mu}\pi)^2\nonumber \\
&&-\frac{1}{2}\mu^2 (\sigma^2+\pi^2)-\frac{\lambda}{4}
(\sigma^2+\pi^2)^2+c\sigma \quad,    
\label{lag}
\end{eqnarray}
where $\mu^2<0$, $c$ is the symmetry-breaking parameter
and $\psi$ is the ``nucleon'' $(N)$ field. 
We take $c\rightarrow 0$ at the final stage.
 
In equal-time quantization the 
NG phase is well described even at the tree-level. 
It is then sufficient 
to demonstrate, by solving the constraints, 
that such a situation is realized
also on LF.  Two kinds of vertices 
will be  examined  
below: the $\sigma \pi\pi$ vertex and the $NN \pi$ vertex. 
\subsection{$\sigma \pi \pi$ {\it vertex}}\par

In this case it is adequate to restrict 
ourselves to the bosonic sector.   
The relevant Lagrangian is 
\begin{equation}
{\cal L}= \frac{1}{2}(\partial_{\mu}\sigma)^2+\frac{1}{2}
(\partial_{\mu}\pi)^2\nonumber \\
-\frac{1}{2}\mu^2 (\sigma^2+\pi^2)-\frac{\lambda}{4}
(\sigma^2+\pi^2)^2+c\sigma \quad.     
\label{blag}
\end{equation}
As in Sect.2, we adopt the periodic boundary condition 
in DLCQ in order to allow the non-vanishing 
vacuum expectation value.  
The quantization can be done in the 
way similar to that in the 
one-component case given in Sect.2.
There are two kinds of
zero modes $\pi_0 
\equiv \frac{1}{2L}\int_{-L}^{L} dx^- \pi(x)$ and 
$\sigma_0 
\equiv \frac{1}{2L}\int_{-L}^{L} dx^- \sigma(x)$
which are separated clearly 
from other oscillating modes,
$\varphi_{\pi}\equiv \pi-\pi_0$ and $\varphi_{\sigma}\equiv 
\sigma-\sigma_0$, respectively.
The canonical commutation relation for the 
oscillating modes 
(\ref{comm}) now reads:
\begin{equation}
\left[\varphi_i(x),\varphi_j(y)\right]
=-{i \over 4}\left\{\epsilon(x^--y^-)-{x^--y^- 
\over L}\right\}\delta_{ij}\delta^{(2)}(x^{\bot}-y^{\bot}) \quad,
\label{commutator}
\end{equation}
where each index $(i)$ stands for $\pi$ or $\sigma$.
By making use of this commutation relation, it is shown that 
the creation and annihilation operators are simply constructed  
from the Fourier coefficients of
$\varphi_{i}$ with respect to $x^{-}$:
\begin{equation}
\varphi_{i}(x)=\sum_{n > 0}\frac{1}{\sqrt{4n \pi}}
\left[a_n^{(i)}(x^+,x^{\bot})e^{-i \frac{\pi n}{L} x^-}
+a _n^{(i) \dagger}(x^+,x^{\bot}) e^{i\frac{n\pi }{L} x^-}\right]
\quad ,  
\label{exp}
\end{equation}
where the coefficients satisfy  
\begin{equation}
[a_n^{(i)}(x^+,x^{\bot}), a_{m}^{\dagger (j)}
(x^+,y^{\bot})]=\delta_{n,  m}
\delta_{i,  j}\delta^{(2)}(x^{\bot}-y^{\bot}) \quad , 
\end{equation}
in spite of the presence of interaction.  
The trivial
Fock vacuum is defined as 
$a_n^{(i)}
\vert 0 \rangle=0$ for any $n$. 
 
Instead of one zero-mode constraint (\ref{zmconstraint2}) 
here we have two coupled zero-mode constraints   
\begin{eqnarray}
\chi_{\pi}&\equiv&
\displaystyle{\frac 1{2L}
\int^L_{-L}dx^-}\left[(\mu^2-\partial_{\bot}^2)\pi
+\lambda \pi(\pi^2+\sigma^2)\right]= 0 \quad , \\
\chi_{\sigma}&\equiv&
\displaystyle{\frac{1}{2L}\int^L_{-L}dx^-}
\left[(\mu^2-\partial_{\bot}^2)\sigma
+\lambda \sigma(\pi^2+\sigma^2)-c \right]=0\quad  , 
\end{eqnarray}
which are also represented by    
\begin{eqnarray}
\chi_{\pi}&=\displaystyle{-\frac{1}{2L}
\int^L_{-L}dx^-2\partial_{+}\partial_{-}\pi=0 } \quad ,\\
\chi_{\sigma}&=\displaystyle{-\frac{1}{2L} 
\int^L_{-L}dx^-2\partial_{+}
\partial_{-}\sigma=0} \quad , 
\end{eqnarray}
through the equation of motion 
(see (\ref{zmconstraint3})).

Our next task is to solve these constraints within some 
approximation. As explained in the beginning of 
this section, it is sufficient  
to obtain a 
solution corresponding to the perturbation theory 
around the classical (tree level) SSB vacuum in equal-time 
quantization.  
For this purpose 
it is convenient to further divide the zero modes as 
\begin{eqnarray}
\pi_{0}&=&v_{\pi}+\omega_{\pi}\quad ,\\
\sigma_{0}&=&v_{\sigma}+\omega_{\sigma}
\quad ,
\end{eqnarray}
where $v_{\pi}$ and $v_{\sigma}$ are the classical constant pieces 
and 
$\omega_{\pi}$ and $\omega_{\sigma}$ are their operator parts. 
Then the zero-mode constraints are split into 
the classical and the operator parts.
The classical parts of the zero-mode constraints are given by 
\begin{eqnarray}
\mu^2v_{\pi}+\lambda v_{\pi}^3+
\lambda v_{\pi}v^2_{\sigma}&=&0 \quad, \\
\mu^2v_{\sigma}+\lambda v_{\sigma}^3+
\lambda v_{\sigma}v^2_{\pi}&=& c 
\quad ,
\end{eqnarray}
which are nothing but the condition determining  
the minima of the classical potential. 
The conventional 
choice of the solution is $v_{\pi}=0$ and 
$v_{\sigma}=v$, where 
$\mu^2v+\lambda v^3=c$. 
 
The operator zero modes 
are solved  
by substituting the perturbative expansion 
\begin{equation}
\omega_i=\sum_{k=1}\lambda^k \omega_i^{(k)}
\quad , 
\end{equation}
into the operator constraints. 
The nontrivial problem which we encounter in solving them is 
the choice of operator ordering, especially
 the ordering between the zero modes and the 
non-zero modes. As mentioned in the Sec 2, 
we {\it assume} the Weyl ordering. In the present context this ordering 
has an 
advantage that  it gives a correct description of 
the symmetric phase 
[$\mu^2>0$, $c\rightarrow 0$] from the view point of 
equal-time quantization. This is closely examined in  
Appendix C. 

The operator part of the zero-mode constraints are 
explicitly written down under the Weyl ordering as follows:
\begin{eqnarray}\label{op-zero2}
(-m_{\pi}^2+\partial_{\bot}^2)\omega_{\pi}
&=&\frac{\lambda}{2L}\int_{-L}^{L}dx^-(\varphi_{\pi}^3
+\varphi_{\pi}\varphi_{\sigma}^2+2v\varphi_{\pi}
\varphi_{\sigma})\nonumber \\ 
&+&\frac{\lambda}{2L}\int_{-L}^{L}dx^{-}
\{(\omega_{\pi}\varphi_{\pi}^2+\varphi_{\pi}^2\omega_{\pi}
+\varphi_{\pi}\omega_{\pi}\varphi_{\pi})
+\frac{1}{2}(\omega_{\pi}\varphi_{\sigma}^2+\varphi_{\sigma}^2
\omega_{\pi}) \nonumber\\
&+&\frac{1}{2}(\omega_{\sigma}\varphi_{\sigma}\varphi_{\pi}
+\varphi_{\pi}\omega_{\sigma}\varphi_{\sigma}
+\varphi_{\sigma}\omega_{\sigma}\varphi_{\pi}
+\varphi_{\sigma}\varphi_{\pi}\omega_{\sigma})\} \\ 
&+&\lambda(\omega_{\pi}^3
+\frac{1}{2}\omega_{\pi}\omega_{\sigma}^2
+\frac{1}{2}\omega_{\sigma}^2\omega_{\pi}
+v\omega_{\pi}\omega_{\sigma}
+v\omega_{\sigma}\omega_{\pi})\quad , \nonumber 
\end{eqnarray}
\begin{eqnarray}\label{op-zero}
(-m_{\sigma}^2+\partial_{\bot}^2)\omega_{\sigma}
&=&\frac{\lambda}{2L}\int_{-L}^{L}dx^{-}(\varphi_{\sigma}^3
+\varphi_{\sigma}\varphi_{\pi}^2+v\varphi_{\pi}^2
+3v\varphi_{\sigma}^2)\nonumber \\ 
&+&\frac{\lambda}{2L}\int_{-L}^{L}dx^{-}
\{(\omega_{\sigma}\varphi_{\sigma}^2+\varphi_{\sigma}^2\omega_{\sigma}
+\varphi_{\sigma}\omega_{\sigma}\varphi_{\sigma})
+\frac{1}{2}(\omega_{\sigma}\varphi_{\pi}^2+\varphi_{\pi}^2
\omega_{\sigma}) \nonumber\\
&+&\frac{1}{2}(\omega_{\pi}\varphi_{\sigma}\varphi_{\pi}
+\varphi_{\pi}\omega_{\pi}\varphi_{\sigma}
+\varphi_{\sigma}\omega_{\pi}\varphi_{\pi}
+\varphi_{\sigma}\varphi_{\pi}\omega_{\pi})\} \\ 
&+&\lambda(\omega_{\sigma}^3
+\frac{1}{2}\omega_{\sigma}\omega_{\pi}^2
+\frac{1}{2}\omega_{\pi}^2\omega_{\sigma}
+v\omega_{\pi}^2
+3v\omega_{\sigma}^2)\quad , \nonumber 
\end{eqnarray}
where each mass term is defined as $m_{\sigma}^2=\mu^2+3\lambda v^2$ 
and $m_{\pi}^2=\mu^2+\lambda v^2$, respectively. 
Here it is worth referring to the scaling property of the zero mode 
with respect to  $x^-$ direction.  
The explicit form of the constraints 
shows that the zero modes are implicitly composed of 
the complicated combination of the following type of integrals 
\begin{equation}
\frac{1}{2L}\int_{-L}^{L}dx^- \varphi_{i}^n(x, L)
\varphi_{j}^m(x, L) \quad,
\label{pol}
\end{equation}
where the explicit $L$ dependence is labeled and $n, m$ 
are some non-negative integers. 
The important feature of (\ref{pol}) is its invariance under the 
scale transformation $L\rightarrow sL$,  
\begin{equation}
\frac{1}{2sL}\int_{-sL}^{sL}dx^- \varphi_{i}^n(x, sL)
\varphi_{j}^m(x, sL)=
\frac{1}{2L}\int_{-L}^{L}dx^- \varphi_{i}^n(x, L)
\varphi_{j}^m(x, L)\quad ,
\end{equation} 
which is on account of the relation  
\begin{equation}
\varphi(sx, sL)=\varphi(x, L)\quad ,
\end{equation}
derived from (\ref{commutator}) or (\ref{exp}). 
Thus  the zero modes have no explicit dependence 
on the box size $L$ and   
the naive continuum limit $L\rightarrow \infty$  
may be safely taken at least for the zero-mode sector. 

Using the explicit form of the zero-mode constraints, 
the lowest order solution of the perturbative zero modes   
for $\omega_{\pi}$ and $\omega_{\sigma}$ is now easily obtained 
\cite{KTY}:
\begin{eqnarray}
(-m_{\pi}^2+\partial_{\bot}^2)\omega_{\pi}^{(1)}
&=&\displaystyle{\frac{\lambda}{2L}\int_{-L}^{L}dx^-\left(\varphi_{\pi}^3
+\varphi_{\pi}\varphi_{\sigma}^2+2v\varphi_{\pi}\varphi_{\sigma}\right)} 
 \quad , \label{pz1}\\ 
(-m_{\sigma}^2+\partial_{\bot}^2)\omega_{\sigma}^{(1)}
&=&\displaystyle{\frac{\lambda}{2L}\int_{-L}^{L}
dx^-\left(\varphi_{\sigma}^3
+\varphi_{\sigma}\varphi_{\pi}^2+v\varphi^2_{\pi}
+3v\varphi^2_{\sigma}\right)}\quad , \label{pz2}
\end{eqnarray}
where  there is no operator-ordering 
ambiguity between the zero modes and the non-zero 
modes. 
Let us briefly see the feature of these explicit solutions. 
One can find that there is no divergence in 
$\omega_{\pi}^{(1)}$ due to the 
positivity  of longitudinal momentum.  
Moreover this feature  is valid 
beyond the leading approximation, that is,  $\omega_{\pi}^{(i)}$ 
has no divergence for any $i\in {\bf N}$, and leads to 
\begin{equation}
\langle \pi \rangle =   \langle \sum_{i=1}^{\infty} 
\omega_{\pi}^{(i)} \rangle +\langle \varphi_{\pi} \rangle= 0 \quad, 
\end{equation}
which is expected from the equal-time perturbation theory.  
We will see that this well-defined zero mode 
solution is used in the actual calculation. 
On the other hand,  $\omega_{\sigma}^{(1)}$ contains the divergence 
which is similar to the tadpole divergence in the equal-time 
perturbation theory. This divergence is essentially the same as the one 
discussed by Robertson \cite{Rob} in the discrete symmetry breaking of 
$\phi^4_2$ model and can be formally renormalized into $v$ through 
the mass renormalization. 

We are now in a position to examine the consequence of the explicit
solutions 
(\ref{pz1}) and (\ref{pz2}).  In order to emphasize 
the importance of $c$, we first 
examine the case $c\equiv 0$ (or $m_{\pi}\equiv0$) again 
 which turns out to have internal inconsistency. 
We will then study the case $c\ne0$ (or $m_{\pi}\ne 0$), 
$c\rightarrow 0$  
and show that the singular behavior of the zero mode (\ref{omega}) 
recovers the correct $\sigma\pi\pi$ vertex \cite{KTY}. 

\noindent
(I) $c\equiv0$ 

The equation of motion  for $\pi$ is given by 
\begin{eqnarray}
\Box\pi(x)&=&j_{\pi}(x) \quad
\nonumber\\
&\equiv&-\lambda\left(\pi^3+\pi\sigma^{\prime
2}+2v\pi\sigma^{\prime}\right) 
\quad,
\end{eqnarray}
where $\sigma^{\prime}=\sigma-v$ and $v=\sqrt{-\mu^2/\lambda}$. 
Rewriting (\ref{emission}) in the present context leads to 
\begin{eqnarray}
(2\pi)^3\delta^{(3)}(\vec{q})
\langle \pi \vert j_{\pi}(0)\vert
\sigma \rangle
\nonumber
&=&\int d^{3}\vec{x} \langle \pi
\vert(2\partial_{+}\partial_{-}-\partial_{\bot}^2)\pi \vert \sigma
\rangle\quad
\label{spp1}\\
&=&-\int d^{3}\vec{x} \langle \pi
\vert \chi_{\pi}(x) \vert \sigma
\rangle=0
\quad ,  \label{spp2}
\end{eqnarray}
where $\vec{q}=\vec{q}_{\sigma}-\vec{p}_{\pi}$ 
is the momentum of the NG boson. 
This relation leads to an internal 
inconsistency: the NG-boson emission 
vertex $\sigma\rightarrow\pi\pi$ at $q^2=0$ 
is vanishing due to the r.h.s. of (\ref{spp1}) or 
equivalently the zero-mode constraint 
$\chi_{\pi}=0$, while the l.h.s. 
of (\ref{spp1}) gives the non-vanishing  
result as will be shown later.  

Furthermore this inconsistency is 
connected with the 
current conservation which also means the charge conservation:
\begin{eqnarray}
\dot{Q}
&=&\frac{d}{dx^+}\int d^3 \vec{y} 
\hspace{2pt}J^+\nonumber\\
                   &=&-\int d^3 \vec{y} \hspace{2pt}
                   (\partial_{-}J^{-}+\partial_{\bot}J^{\bot}) 
                   \nonumber\\
                    &=&-\int d^2 y^{\bot} 
                   \{J^{-}(x^-=+L)-J^{-}(x^-=-L)\} =0 \quad , 
\label{cc}                   
\end{eqnarray}  
where the periodic boundary condition 
was used. In our model we have 
an explicit form of the O(2) current
\begin{equation}
 J^{\mu}=\partial_{\mu}\sigma\pi-\partial_{\mu}\pi
\sigma \quad.
\label{current}
\end{equation}
The LF charge defined by $Q=\int d^3\vec{x} J^+$ reduces to 
\begin{equation}
Q=\int d^3\vec{x}
(\partial_{-}\varphi_{\sigma}\varphi_{\pi}-\partial_{-}
\varphi_{\pi}\varphi_{\sigma}) \quad , 
\label{chargenozm}
\end{equation}
which contains only the oscillating modes,  
because the operator part of the zero mode in addition to the 
pole term is 
dropped by the integration over $x^-$
\cite{MY}.  
Thus the LF charge is well-defined even in the NG phase and always 
annihilates the vacuum simply by the $P^+$ conservation: 
\begin{equation}
Q \vert  0 \rangle=0 \quad, 
\label{vac-annih}
\end{equation}    
which supports the trivial property of the LF vacuum. 
This will also be 
checked in later discussions, 
see (\ref{trivvac0}) and (\ref{trivvac}).

The charge conservation can also be checked through the explicit
calculation 
including  the zero modes. 
In fact the straightforward but tedious calculation using the
perturbative 
solution 
of the zero modes leads to  
\begin{equation}
[Q, P^-]=i\int d^3 \vec{x} 
(v\chi_{\pi}+\omega_{\sigma}\chi_{\pi}-\omega_{\pi}\chi_{\sigma})
+(divergence)
\quad,
\label{Q-H}
\end{equation}
where the divergence arises from 
the operator ordering and should be 
renormalized in an appropriate way. 
If we simply neglect this  
divergence, or the commutator is understood as Dirac bracket, 
the r.h.s. of (\ref{Q-H}) becomes zero owing 
to the zero-mode constraints. 

By substituting $\sigma=\sigma^{\prime}+v$ 
into (\ref{current}), we have   
\begin{equation}
        J^{\mu}=\widehat J^{\mu}-v\partial_{\mu}\pi\quad, 
\end{equation}
where $\widehat
J^{\mu}=\partial_{\mu}\sigma^{\prime}\pi-\partial_{\mu}\pi
\sigma^{\prime}$. 
 From the current conservation, we obtain 
\begin{eqnarray} 
\langle \pi \vert 
\int d^3\vec{x}\, \partial_{\mu}\widehat J^{\mu}(x)\vert \sigma
\rangle_{x^+=0}
&=&-v\langle \pi \vert 
\int d^3\vec{x}\chi_{\pi} \vert \sigma
\rangle_{x^+=0}
\nonumber
\\
&=&-i(2\pi)^3\delta^{(3)}(\vec{q})\, \displaystyle{
\frac{m_{\sigma}^2-m_{\pi}^2}{2p_{\sigma}^+}}\langle 
\pi \vert \widehat J^+(0)
\vert \sigma \rangle \quad ,  
\end{eqnarray} 
which means $\langle \pi \vert \widehat J^+(0)
\vert \sigma \rangle=0$ 
($m_{\sigma}^2\ne m_{\pi}^2$) 
due to the charge conservation or the zero-mode 
constraint.  
The charge conservation does not 
lead to the NG phase at all.  

This pathology 
suggests that the zero-mode constraint without mass term 
is ill-defined. Up to the operator ordering, 
the zero-mode constraint for $\omega_{\pi}$ with $m_{\pi}\equiv0$ is
given  by 
\begin{eqnarray}
\partial_{\bot}^2\omega_{\pi}
&=&\frac{\lambda}{2L}\int_{-L}^{L}dx^-(\varphi_{\pi}^3
+\varphi_{\pi}\varphi_{\sigma}^2+2v\varphi_{\pi}
\varphi_{\sigma})\nonumber \\ 
&+&\frac{\lambda}{2L}\int_{-L}^{L}dx^{-}
(\omega_{\pi}\varphi_{\sigma}^2+2\omega_{\sigma}
\varphi_{\sigma}\varphi_{\pi}
+3\omega_{\pi}\varphi_{\pi}^2)\nonumber \\
&+&\lambda(\omega_{\pi}^3
+\omega_{\pi}\omega_{\sigma}^2
+2v\omega_{\pi}\omega_{\sigma}) \quad . 
\label{piz}
\end{eqnarray} 
Note that it is not $\omega_{\pi}$ but $
\int d^3\vec{x} \hspace{2pt}\omega_{\pi}$ which is used 
to calculate the $\sigma\pi\pi$ vertex.  
Thus the real quantity to be considered is the integration of
(\ref{piz}) 
over the LF which 
has the similar structure as that in two dimensions, because 
the l.h.s. of (\ref{piz}) vanishes by the transverse integration. 
However, in two dimensions,
the zero-mode constraint without 
mass term is ill-defined in the interacting theory. 
The reason can be easily understood as follows. The  
overall factor $\lambda$ appearing on the r.h.s. of 
(\ref{piz}) is canceled and dropped out.  
The solution $\omega_{\pi}$ is  
then independent of $\lambda$. The Eq. (\ref{piz}) (before the
transverse 
integration) dictates that $\omega_{\pi}$ does depend on 
$\lambda$.   
Therefore  
it is  necessary to 
introduce the NG-boson mass to make the theory well-defined.  

\noindent
(II) $c\ne0$ {\it and} $c\rightarrow 0$

Having the explicit breaking NG boson mass, 
we now derive the tree-level matrix element 
for the $\sigma\pi\pi$ scattering \cite{KTY}.  
The singular behavior of the NG-boson zero mode proposed in the 
previous section is in fact derived from (\ref{pz1}):
\begin{equation}
\lim_{m_{\pi}^2\rightarrow 0} m_{\pi}^2\int d^3 \vec{x}\,
\omega_{\pi}^{(1)}
=-\lambda\int d^3 \vec{x}\, (\varphi_{\pi}^3
+\varphi_{\pi}\varphi_{\sigma}^2+2v\varphi_{\pi}\varphi_{\sigma})
\ne 0\quad,  
\label{omega2}
\end{equation}
which is not restricted to the lowest order but is valid for higher
order 
solutions:
\begin{equation}
\int d^3 \vec{x}\omega_{\pi}^{(n)}\sim \frac{1}{m_{\pi}^2}
\quad . 
\label{omega3}
\end{equation}
See Appendix D for more details. 

This actually ensures the correct 
$\sigma \rightarrow \pi \pi$ vertex 
as well as the non-vanishing current vertex. 
Indeed  the $\sigma \rightarrow \pi \pi$ vertex  
at $q^2=0$ is evaluated as follows:
\begin{eqnarray}
\langle \pi,\pi(q^2=0)\vert 
\sigma\rangle&=&\lim_{m^2_{\pi}\rightarrow 0}i\int d^{4}x \hspace{2pt}
e^{iq x}\langle \pi\vert (\Box+m_{\pi}^2)\pi(x)\vert 
\sigma \rangle \quad \nonumber\\
&=&\lim_{m^2_{\pi}\rightarrow 0}i\int dx^{+}\hspace{2pt} 
e^{iq^- x^+}\langle \pi\vert m_{\pi}^2\int_{-\infty}^{+\infty}d^3
\vec{x}
\hspace{5pt}\omega_{\pi}^{(1)}\vert \sigma \rangle \quad \label{vertex}
\nonumber\\
&=&i(2\pi)^4\delta(q^-+p^-_{\pi}-p^-_{\sigma})
\delta^{(3)}(\vec{p}_{\sigma}
-\vec{p}_{\pi})(-2\lambda v) \quad,
\end{eqnarray} 
where $q^{\mu}=p^{\mu}_{\sigma}-p^{\mu}_{\pi}$ is the 
four momentum of the NG boson.  
The current vertex is also obtained by using 
the operator relation
\begin{equation}
\partial_{\mu}J^{\mu}(x)=
vm_{\pi}^2\pi(x)\quad,
\end{equation} 
which leads to 
\begin{equation}
\partial_{\mu}\widehat J^{\mu}(x)=
v(\Box+m_{\pi}^2)\pi(x)=vj_{\pi}(x)\quad, 
\end{equation}
where 
$j_{\pi}(x)
=-\lambda(\pi^3+\pi\sigma^{\prime 2}+2v\pi\sigma^{\prime})$ with 
$\sigma^{\prime}=\sigma-v$. Then we can confirm the following 
relations: 
\begin{eqnarray}\label{cver}
\langle\pi\vert \int d^3\vec{x}
\partial_{\mu}\widehat J^{\mu}(x)
\vert\sigma\rangle
&=&\int d^3\vec{x}
\langle \pi \vert vj_{\pi}(x)\vert\sigma\rangle \nonumber \\ 
&=&v m^{2}_{\pi}\langle\pi\vert
\int d^3\vec{x} \omega_{\pi}^{(1)}(x)\vert\sigma\rangle \nonumber \\
&=&-2\lambda v^{2}(2\pi)^{3}\delta(q^{+})
\delta^{(2)}(q^{\bot})
\end{eqnarray}
in the limit of $m_{\pi}\rightarrow 0$ 
or $v\rightarrow \sqrt{\frac{-\mu^2}{\lambda}}$.
Throughout the calculations we have 
used the covariant normalization of states 
$\langle p_{\beta}\vert p_{\alpha}\rangle=
(2\pi)^3 2p^+_{\alpha}\delta^{(3)}(\vec{p}_{\alpha}-
\vec{p}_{\beta})$ and the on-shell mode expansion for $\varphi$ 
in the continuum limit.

Let us see the property of the LF charge in this case ($c\ne 0$). 
As already mentioned, 
the LF charge is well-defined even in the NG phase and always 
annihilates the vacuum simply by the $P^+$ conservation: 
\begin{equation}
Q \vert  0 \rangle=0 \quad, 
\end{equation}    
which supports the trivial property of the LF vacuum. 
The trivial LF vacuum is also consistent with the explicit 
computation of the commutators. We can show that the oscillating  
modes are  transformed under the action of $Q$ as  
\begin{equation}
 [Q, \varphi_{\sigma}] =-i \varphi_{\pi} \quad \mbox{and} \quad 
 [Q,\varphi_{\pi}] = i\varphi_{\sigma}\quad, 
\label{osctrans}
\end{equation}
where use has been made of (\ref{commutator}) and (\ref{chargenozm}).
Then we have 
\begin{equation}
\langle[Q, 
\varphi_{\sigma}]\rangle=\langle[Q,\varphi_{\pi}]\rangle=0 \quad,
\label{trivvac0}
\end{equation}
 which is consistent with the trivial vacuum: $Q\vert 0\rangle=0$. 
On the other hand, the lowest order solution of the zero modes yields
rather complicated transformation property of the zero modes:  
\begin{eqnarray}
\left[Q, \omega_{\sigma}\right]
&=&(-m_{\sigma}^2+\partial_{\bot}^2)^{-1}
\displaystyle{\frac{i\lambda}{2L} \int_{-L}^{L}dx^-( \varphi_{\pi}^3
+\varphi_{\pi}\varphi_{\sigma}^2+4v\varphi_{\pi}
\varphi_{\sigma})\ne-i\omega_{\pi}}\quad,
\nonumber \quad \\
\left[ Q, \omega_{\pi} \right] & = & (-m_{\pi}^2+\partial_{\bot}^2)^{-1}
\displaystyle{\frac{i\lambda}{2L} \int_{-L}^{L}dx^-(\varphi_{\sigma}^3
+\varphi_{\sigma}\varphi_{\pi}^2-2v\varphi_{\pi}^2
+2v\varphi_{\sigma}^2)}\ne i\omega_{\sigma}\quad. 
\label{zmtrans}
\end{eqnarray}
Nevertheless, it is straightforward to confirm that 
\begin{equation}
\langle [Q, \pi]\rangle=\langle [Q, \sigma]\rangle=0
\label{trivvac}
\end{equation}
by putting together (\ref{osctrans}) and (\ref{zmtrans}).
Eq.(\ref{trivvac}) can also be checked by use of the commutator 
for the full fields (\ref{comm1}) or (\ref{comm2}) 
up to operator ordering: 
Would-be non-vanishing term arising
from the sign function is precisely cancelled by that from
the extra term in the commutator. 
Then we conclude that our LF charge does annihilate
the vacuum in accord with the general 
argument \cite{MY} for the trivial vacuum 
based on the zero-mode constraint through 
which the zero mode can be solved away 
out of the physical Fock space.
 We should remark that $[Q, \pi]\ne i\sigma$ 
and $[Q, \sigma]\ne -i\pi$ 
even in the $m_{\pi}\rightarrow 0$ limit which are 
on account of the effect of zero modes. They 
are contrasted to those in the usual equal-time case: $[Q^{\rm 
et},\sigma]=-i\pi$, and 
$[Q^{\rm et},\pi]=i\sigma$. Since the information of the equal-time 
vacuum symbolically denoted by 
``$Q^{\rm et}\vert 0\rangle_{\rm et}\ne 0$'' is 
expected to be carried into the zero mode in DLCQ,  
the unusual transformation property 
of the zero modes in the NG phase seems to be natural.  

Finally, we can show from (\ref{cver}) that the regularized 
zero mode leads to non-conservation of the LF charge in the
symmetric limit of $m_{\pi}^2\rightarrow 0$ \cite{KTY}:
\begin{equation} 
\dot{Q}=\frac{1}{i}[Q, P^-]=v 
\lim_{m_{\pi}^2\rightarrow 0} m_{\pi}^2
\int d^3\vec{x}\, \omega_{\pi}\neq 0\quad.
\end{equation}
Therefore the SSB in DLCQ 
is realized as if it were an 
explicit symmetry breaking. 
Actually, there exists no NG theorem on the LF. Instead,
the singular behavior (\ref{omega3}) establishes existence of the
massless 
NG boson coupled to
the current whose charge satisfies $Q\vert 0\rangle=0$ 
and $\dot{Q}\ne 0$, 
in much the same as the NG theorem in the equal-time 
quantization which
ensures existence of the massless NG boson coupled to the current 
whose charge satisfies $Q^{\rm et}\vert 0\rangle_{\rm et}
 \ne 0$ and $\dot{Q^{\rm et}} =0$.

\subsection{$NN\pi$ {\it vertex}}\par
As was shown in the previous subsection, 
the classical part of the zero mode  
for $\sigma$ is given by $v$ which contributes to 
the vacuum expectation 
value $\langle\sigma\rangle=v$ in the trivial LF vacuum. 
Rewriting the Lagrangian (\ref{lag}) by the shifted 
field $\sigma^{\prime}=\sigma-v$, 
we find the standard Lagrangian in the broken phase
\begin{eqnarray}
{\cal L}&=&\bar{\psi}\{i\gamma^{\mu}\partial_{\mu}-M(x)\}\psi
+ \frac{1}{2}(\partial_{\mu}\sigma^{\prime})^2+\frac{1}{2}
(\partial_{\mu}\pi)^2\nonumber \\
&-&\frac{1}{2}m_{\sigma}^2 \sigma^2-\frac{1}{2}m_{\pi}^2\pi^2
-\frac{\lambda}{4}(\sigma^{\prime})^4
-\frac{\lambda}{4}(\pi)^4-\lambda v(\sigma^{\prime})^3
-\lambda v\sigma^{\prime}\pi^2
-\frac{\lambda}{2}(\sigma^{\prime})^2\pi^2\quad,    
\label{lag2}
\end{eqnarray}
where $M(x)=m_{\psi}+g(\sigma^{\prime}(x)+i\gamma_{5}\pi(x))$ and 
the ``nucleon'' field $\psi$ 
acquired the ``degenerate'' mass $m_{\psi}=gv$. 

Let us first clarify the fermion contribution 
to the NG-boson zero mode. 
By integrating the equation of motion over $x^-$,  
\begin{equation}
(-2\partial_{+}\partial_{-}+\partial_{\bot}^2-m_{\pi}^2)\pi
-\lambda\{\pi^3
+\pi(\sigma^{\prime})^2+2v\sigma^{\prime}\pi\}
-g\bar{\psi}i\gamma_{5}\psi=0 \quad , 
\end{equation}
we can easily derive 
the zero-mode constraint for the NG boson:
\begin{equation}
(-m_{\pi}^2+\partial^2_{\bot})\omega_{\pi}
=\frac{g}{2L}\int_{-L}^L dx^- \bar{\psi}i\gamma_{5}\psi
+(\mbox{scalar and pseudo-scalar parts}) \quad.    
\label{f-zero}
\end{equation} 
Henceforth we 
shall omit both scalar and pseudo-scalar parts for 
simplicity. 

As we have seen in  the $\sigma\pi\pi$ vertex, 
the $NN\pi$ vertex at $q^2=0$ is essentially  given by 
\begin{equation}
-m^{2}_{\pi}\int d^3\vec{x} \hspace{2pt}\omega_{\pi}=
g\int d^3\vec{x} \hspace{2pt}\bar{\psi}i\gamma_{5}\psi \quad,  
\label{psd}
\end{equation}
which is consistent to our proposal  (\ref{omega}) 
as long as the r.h.s. is non-vanishing. Let us 
estimate the r.h.s. of (\ref{psd}) in detail.
We assume the anti-periodic boundary condition 
for the fermion field and neglect the fermion zero mode.  
Introducing the projection operator 
$\Lambda_{(\pm)}=\frac{1}{\sqrt{2}}\gamma^0\gamma^{\pm}$
where $\gamma^{\pm}=\frac{1}{\sqrt{2}}(\gamma^{0} \pm \gamma^{3})$,  
the fermion field can be decomposed into the dynamical plus the 
non-dynamical components $\psi=\psi_{(+)}+\psi_{(-)}$ where 
$\psi_{(\pm)}=\Lambda_{(\pm)}\psi$. 
Note that $(\gamma^+)^2=(\gamma^-)^2=0$ implies 
$\Lambda_{(+)}\Lambda_{(-)}=\Lambda_{(-)}\Lambda_{(+)}=0$.   
The relevant part of the  Lagrangian  
can be  written in terms of the two kinds of fermion projections: 
\begin{eqnarray}
\cal{L}&=&\bar{\psi}(x)\{i\gamma^{\mu}\partial_{\mu}-M(x)\}\psi(x) 
\nonumber \\
       &=&i \sqrt{2}\psi^{\dagger}_{(+)}\partial_{+}\psi_{(+)}
          +i \sqrt {2}\psi^{\dagger}_{(-)}\partial_{-}\psi_{(-)}
          \quad  \nonumber\\
          &+&\psi^{\dagger}_{(+)}(i\gamma^0\gamma^k\partial_k 
            -\gamma^0M)\psi_{(-)}
            +\psi^{\dagger}_{(-)}(i\gamma^0\gamma^k\partial_k 
            -\gamma^0M)\psi_{(+)}\quad. 
\label{bosonic}
\end{eqnarray}
The equation of motion for  $\psi_{(-)}$ leads to the 
constraint equation 
\begin{equation}
i\sqrt{2}\partial_{-}\psi_{(-)}+(i\gamma^0\gamma^k\partial_k 
            -\gamma^0M)\psi_{(+)}=0 \quad ,  
\end{equation} 
whose solution is 
readily obtained as 
\begin{equation}
\psi_{(-)}(x)=\frac{i}{2\sqrt{2}}\int_{-L}^L dy^{-}
            \epsilon(x^- -y^-)\{i\gamma^0\gamma^k\partial_k 
            -\gamma^0M(y)\}\psi_{(+)}(y) \quad. 
\label{p-sol}
\end{equation}
By substituting the
solution of $\psi_{(-)}$  into 
 the r.h.s. of (\ref{psd}), one 
obtains 
\begin{equation}
\int d^3\vec{x}\hspace{2pt}\bar{\psi}i\gamma_{5}\psi 
              =\frac{1}{2\sqrt{2}}\int
d^3\vec{x}dy^{-}\epsilon(x^{-}-y^{-})
     \psi^{\dagger}_{(+)}(x)\{M(x)\gamma_5-\gamma_5 M(y)\} 
     \psi_{(+)}(y)\quad,  
     \label{psd2}
\end{equation}
where the r.h.s. is non-vanishing as is seen from the explicit 
form of $M(x)=m_{\psi}+g\sigma^{\prime}(x)+ig\gamma_{5}\pi(x)$.  
Thus our  proposal for the singular behavior of the NG-boson zero 
mode is confirmed.
 
It is worth mentioning that exceptional situation 
happens for the tree-level $NN\pi$ coupling, in which case  
we should keep only the 
$x-$independent 
part of 
$M(x)$, i.e., $m_{\psi}$ in (\ref{psd2}), because the rest gives 
higher order contributions. 
Keeping only $m_{\psi}$ is equivalent to 
applying
the free equation of motion 
to the r.h.s. of (\ref{psd}).  
It is a well-known peculiarity on the LF that  
the pseudo-scalar density for {\it free massive} fermion, when 
integrated over the LF, vanishes.  This  is what the r.h.s. of 
(\ref{psd2})  
implies for $g=0$ and $M=m_{\psi}$. Owing to this specific 
property on the LF,   
the physical amplitude for the tree-level $NN\pi$ scattering 
at $q^2=0$ becomes zero,  which is of course 
consistent with that in the equal-time 
quantization. Note that  
$q^2=0$ is nothing but the soft momentum limit $q^{\mu}=0$ for the 
``degenerate nucleon mass'' 
and the physical amplitude 
of the NG-boson emission vertex as well as the associated  
current vertex is known to be zero in such a limit  
for kinematical reason from the low energy theorem,
even when $NN\pi$ coupling constant and $g_A$ are non-zero \cite{BKY}.  


\section{Zero-Mode Problem in the Continuum LF Quantization}\par
\my

The issue of symmetry breakings is important not only in the 
DLCQ but also in the continuum LF framework such as 
the renormalization-group approach. 
Wilson et al. \cite{Wils} studied the sigma model 
``without zero mode" in the continuum framework  
and described the broken phase at the tree level 
by constructing the corresponding ``effective Hamiltonian" 
{\it without zero mode} 
and with the ``unusual counter terms" which 
compensate the ``removal of the zero mode". 
Instead of comparing  our  result with theirs in a direct manner,  
we here examine the same sigma model in the broken 
phase in the general continuum framework, 
{\it paying special attention to 
the boundary condition}. As we emphasized 
in Section 2 (also in Appendix B),
the boundary condition in the LF quantization 
contains dynamical information 
and is crucial to define the theory.
Then we shall demonstrate that it is actually {\it 
impossible to remove the zero mode
in the continuum theory} in a manner 
consistent with the trivial vacuum. 
The point is that the real problem with the zero mode in the
continuum theory is {\it not a single mode} 
with $p^+ \equiv 0$, which is just measure zero, 
but the {\it accumulating point} 
$p^+ \rightarrow 0$ \cite{NY}.
This is in sharp contrast to our result in DLCQ where 
the trivial vacuum 
is always guaranteed thanks to the clean separation 
and explicit removal of 
the zero mode through the zero-mode constraint.

Let us illustrate this by starting with the canonical commutator
for the fields $\sigma,\pi$ in the bosonic part of the 
$O(2)$ sigma model (\ref{blag}) 
(without explicit symmetry breaking term, $c\equiv 0$)
 {\it in the continuum theory}:
\begin{equation}
[\sigma(x), \sigma(y)]_{x^+=y^+}=-\frac{i}{4}\epsilon(x^--y^-)
\delta^{(2)}(x^{\perp}-y^{\perp})\quad, 
\label{FSc}
\end{equation}
\begin{equation}
[\pi(x), \pi(y)]_{x^+=y^+}=-\frac{i}{4}\epsilon(x^--y^-)
\delta^{(2)}(x^{\perp}-y^{\perp})\quad, 
\label{FScc}
\end{equation}
where the sign function
\begin{equation}
\epsilon (x^-) =\frac{i}{\pi} {\cal P}
\int_{-\infty}^{+\infty}\frac{dp^+}{p^+}
e^{-ip^+x^-}
\end{equation}
is defined by the principal value prescription 
and hence has no $p^+\equiv 0$ mode but does have an
accumulating point $p^+ \rightarrow 0$. 
This accumulating point is really
a trouble as we will see in the followings.
Then, {\it as far as we use this sign function}
for the commutator, {\it we cannot really remove the zero mode} 
in this sense.

We first look at the transformation property
of the fields $\sigma, \pi$.
The conserved current associated with the symmetry of the Lagrangian 
is given by 
\begin{equation}
J_{\mu}=\pi\partial_{\mu}\sigma-\sigma\partial_{\mu}\pi\quad,   
\label{FSa}
\end{equation}  
and the  LF charge is defined by 
\begin{equation}
Q=\int d^3 \vec{x}( \pi\partial_{-}\sigma-\sigma\partial_{-}\pi) 
\quad. 
\label{FSb}
\end{equation}
 From the canonical commutation relations (\ref{FSc}) and (\ref{FScc})   
we can easily find 
\begin{eqnarray}
&&[Q, \sigma(x)]=-i\pi(x)+\frac{i}{4}[\pi(x^-=\infty)
+\pi(x^-=-\infty)] \quad,\\
&&[Q, \pi(x)]= i\sigma(x)-\frac{i}{4}[ \sigma(x^-=\infty)
+\sigma(x^-=-\infty)] \quad.
\end{eqnarray}

To obtain a
sensible transformation property of the 
fundamental fields, the surface terms must 
vanish as operators:
\begin{equation}
\pi(x^-=\infty)+\pi(x^-=-\infty)
=\sigma(x^-=\infty)+\sigma(x^-=-\infty)=0 \quad.   
\label{FSd}
\end{equation}
However, this condition, anti-periodic boundary condition,
means that the zero mode is not allowed to exist 
and hence its classical part, condensate $\langle \sigma \rangle$, 
does not exist at all. 
Thus we have no spontaneous symmetry breaking contrary to 
the initial assumption. 

We then seek for a modification of the boundary condition
to save the condensate and vanishing surface term simultaneously.
The lesson from the above argument is that 
we cannot impose the canonical commutation relation  
for the full fields, because then not only the surface term but also 
the zero mode (and hence condensate) are required to vanish 
due to the relation (\ref{FSd}). 
So,  let us 
first separate the constant part or condensate (classical zero mode)
$v$ from $\sigma$
and then impose the canonical commutation relations for the 
fields without zero modes, 
$\pi$ and the shifted field $\phi=\sigma-v=
\sigma^{\prime}$ (here we use $\phi$ instead of $\sigma^{\prime}$),  
which are now consistent with the anti-periodic boundary
condition and (\ref{FSd}). This actually corresponds to the usual
quantization
around the classical SSB vacuum in the equal-time quantization.  
The constant part $v$ should be understood to be determined 
by the 
minimum of the classical potential
\begin{eqnarray}
V&=&\frac{1}{2}\mu^2(\sigma^2+\pi^2)
        +\frac{\lambda}{4}(\sigma^2+\pi^2)^2 \quad,\nonumber \\
 &=&\frac{1}{2}m_{\phi}^2\phi^2+\lambda v\phi(\phi^2+\pi^2)
 +\frac{\lambda}{4}(\phi^2+\pi^2)^2\quad,
\label{FSe}
\end{eqnarray}   
where $v=\sqrt{-\mu^2/\lambda}$, $\mu^2<0$ and 
$m_{\phi}^2=
2\lambda v^2$. In the renormalization group approach, the potential  
(\ref{FSe}) appears as an ``effective Hamiltonian" \cite{Wils},
while the same potential can be obtained simply through shifting
$\sigma$ 
to $\phi=\sigma-v$.  
The canonical commutation relation for $\sigma$ 
is now replaced by 
\begin{equation}
[\phi(x), \phi(y)]_{x^+=y^+}=-\frac{i}{4}\epsilon(x^--y^-)
\delta^{(2)}(x^{\perp}-y^{\perp})\quad.   
\label{FSf}
\end{equation}

Now that the quantized
fields have been arranged 
to obey the anti-periodic boundary condition, 
one might consider that we have removed the zero mode.
It is not true, however, {\it as far as we are using the 
commutator with the sign function}, (\ref{FSf}), 
in which the zero mode as an accumulating point
is persistent to exist.

Let us look at the LF charge which is given by 
\begin{equation}
Q=\int d^3 \vec{x}( \pi\partial_{-}\phi-\phi\partial_{-}\pi-v\partial
_{-}\pi) \quad. 
\label{FSg}
\end{equation}
The straightforward calculation leads to 
\begin{equation}
[Q, \phi(x)]=-i\pi(x)+\frac{i}{4}[\pi(x^- =\infty)+\pi(x^- =-\infty)]
\quad,
\label{crphis}
\end{equation}
\begin{equation}
[Q, \pi(x)]=i\phi(x)+\frac{i}{2}v-
\frac{i}{4}[\phi(x^- =\infty)+\phi(x^- =-\infty)] \quad,  
\label{crpis}
\end{equation}
where the surface terms should vanish:   
\begin{equation}
\phi(x^- =\infty)+\phi(x^- =-\infty)=\pi(x^- =\infty)+\pi(x^-
=-\infty)=0 \quad 
\label{FSh}
\end{equation}
for the same reason as before. Thus we find   
\begin{equation}
[Q, \phi(x)]=-i\pi(x)\quad,
\end{equation}
\begin{equation}
[Q, \pi(x)]=i\phi(x)+\frac{i}{2}v \quad. 
\label{FSi1}
\end{equation}
The constant term on the r.h.s. of (\ref{FSi1}) has its origin in the 
commutation relation (\ref{FScc}), or equivalently,
\begin{equation}
[\partial_{-}\pi(x), \pi(y)]=-\frac{i}{2}\delta^{(3)}(\vec{x}-\vec{y}) 
\quad.  
\label{FSccc}
\end{equation} 
Namely, 
\begin{eqnarray}
\langle 0 \vert [Q, \pi(x)]\vert 0 \rangle&=&i\langle 0 \vert 
\phi(x)\vert 0 \rangle+\frac{i}{2}v \quad, 
\nonumber\\
 &=& \frac{i}{2}v\ne 0 \quad. 
\end{eqnarray}
Then we find that the {\it LF charge does not annihilate the vacuum} 
$Q\vert 0 \rangle \ne 0$
and we have lost 
the trivial vacuum which is a vital feature of the LF quantization. 
There actually exist infinite number of zero-mode states 
$\vert \alpha \rangle\equiv e^{i\alpha Q}\vert 0\rangle$
such that $P^+\vert \alpha\rangle = e^{i\alpha Q} P^+\vert 0\rangle=0$,
where we have used
$[P^+,Q]=0$ and $\alpha$ is a real number: All these states satisfy the 
``Fock-vacuum condition'' $a(p^+)\vert \alpha\rangle =0$ and 
hence the true unique vacuum cannot be specified by this condition in
contrast to the usual expectation.
This implies that the 
zero mode has not been removed, 
even though the Hamiltonian has been rearranged 
by shifting the field into the one without exact zero mode $p^+\equiv
0$.
This is in sharp contrast to DLCQ in Sect.4 
where the surface terms in (\ref{crphis}) and (\ref{crpis}) 
and the constant term in (\ref{FSi1}) do vanish 
altogether thanks to the additional term
$-(x^- - y^-)/L$  (``subtraction of the zero mode") besides the sign
function 
$\epsilon (x^- - y^-)$ in the canonical commutator (\ref{commutator}).

It should be noted that somewhat peculiar situation happens to the LF
charge due to this boundary condition at the surface term: 
Although the local current is conserved, the LF charge is not. 
In fact, integrating the equation of the current conservation
$\partial_{\mu}J^{\mu}=0$,
we have 
\begin{eqnarray} 
\frac{dQ}{dx^+}&=&-\int d^3 \vec{x}(\partial_{-}J^-
+\partial_{\perp}J^{\perp})\quad\nonumber\\  
      &=&v\int d^3 \vec{x}\partial_{+}\partial_{-}\pi \quad\nonumber\\
               &=&v\int d^2 x^{\perp}[\partial_{+}\pi(x^-=\infty)
               -\partial_{+}\pi(x^-=-\infty)]\quad \nonumber\\
               &=&2v\int d^2 x^{\perp}\partial_{+}\pi(x^-=\infty)\ne 0
\quad, 
\end{eqnarray}
where the anti-periodic boundary condition (\ref{FSh}) has been used.
Thus in the continuum theory {\it LF charge is not conserved} 
in spite of the conservation of the local current.
This charge non-conservation can also be checked by direct calculation: 
\begin{eqnarray}
i\frac{dQ}{dx^+}&=&[Q, H] \quad \nonumber\\
                &=&-\frac{i}{2}\lambda v\int d^3 \vec{x} 
                [\pi(\phi^2+\pi^2)+2v\phi\pi] \quad \nonumber\\
                &=&iv\int d^2 x^{\perp}[\partial_{+}\pi(x^-=\infty)
               -\partial_{+}\pi(x^-=-\infty)]\quad, 
\label{FSj}
\end{eqnarray}
where use has been made of the equation of motion 
\begin{equation}
-2\partial_{+}\partial_{-}\pi+\partial_{\perp}^{2}\pi=2\lambda v \phi
\pi 
+\lambda\pi(\phi^2+\pi^2) \quad
\end{equation}
as well as the anti-periodic boundary condition (\ref{FSh}). 

The resulting Hamiltonian via the field shifting 
coincides with the ``effective
Hamiltonian" of Ref.\cite{Wils} which was obtained 
by ``removing the zero mode and
adding unusual counter terms" for it. The above peculiarity of the LF
charge,
non-conservation of the LF charge and conservation of the local current, 
was also claimed in Ref.\cite{Wils} for completely different reason than
ours.
They implicitly assumed vanishing surface terms altogether:
\begin{equation}
\phi(x^- =\infty)=\phi(x^- =-\infty)=\pi(x^- =\infty)=\pi(x^-
=-\infty)=0 \quad. 
\label{FSi}
\end{equation}
However, it is actually not allowed, 
because it contradicts the commutation 
relation (\ref{FScc}) and (\ref{FSf}). For instance, 
the commutation relation (\ref{FSccc}) yields
\begin{equation}
\big[ \int d^3 \vec{x} \partial_{-}\pi(x), \pi(y) \big]_{x^+=y^+} 
=\int d^3 \vec{x} \big[\partial_{-}\pi(x), \pi(y) \big]_{x^+=y^+} 
 =-\frac{i}{2}\ne 0\quad,
\end{equation}
while (\ref{FSi}) requires l.h.s. to vanish.
If one illegitimately assumed  (\ref{FSi}) and neglected 
all surface terms, then the LF charge would have been conserved 
as is easily read off from (\ref{FSj}), 
in contradiction to \cite{Wils}.  
 
To summarize, in the general continuum LF quantization based on the
canonical 
commutation relation {\it with sign function}, 
the LF charge does {\it not annihilate}
 the vacuum and is {\it not conserved} for the 
conserved local current. 
It corresponds to impossibility to remove the zero
mode as an accumulating point in the continuum theory in a manner
consistent 
with the trivial vacuum. Thus, in the continuum theory 
the greatest advantage of the
 LF quantization, the simplicity of the vacuum, is lost, although the 
 NG-boson emission vertex can be nontrivial without such 
a manipulation as via the 
 explicit NG boson mass in contrast to DLCQ.  

Here it is worth suggesting that even in the continuum theory there
exists a
prescription, ``$\nu$-theory'' \cite{NY}, which may
give rise to the same result as that we obtained in DLCQ in this paper:
the trivial vacuum $Q\vert 0 \rangle=0$, and the 
non-decoupling NG boson through the
explicit breaking mass of the NG-boson 
or the singular behavior of the global zero mode
of the NG boson. The $\nu$-theory modifies the sign function in the
commutator into a certain function which vanishes at $x^-=\pm \infty$,
by shaving the vicinity of the zero mode to tame the $1/p^+$ singularity
as $|p^+|^{\nu}/p^+ (\nu >0)$. 
The limit of $\nu\rightarrow 0$ is taken only after whole calculation. 
Then there 
is no surface term nor
constant term ($\frac{i}{2}v$) in the commutator (\ref{crphis}) and 
(\ref{crpis}),
and hence the transformation property of the fields and the trivial
vacuum 
should be both realized. Also, the LF charge conservation is expected to
follow unless we introduce the explicit symmetry
breaking, which is in fact the same situation as in DLCQ. Thus, 
in order to realize the NG phase we could do the same game as DLCQ,
namely, 
introduce the explicit breaking mass of the 
NG boson in such a way that the global zero mode of the NG boson 
behaves singular as $1/m_{\pi}^2$ in the symmetric limit.


\section{Summary and Discussions} 
\my

We have studied how the continuous symmetry breaking in (3+1)-dimensions
is described on the LF within the framework of DLCQ.  We have shown that 
it is necessary to introduce an 
explicit
symmetry-breaking mass 
of the NG boson $m_{\pi}$ in order to realize the NG phase in DLCQ. 
The NG phase is reproduced in the limit of  $m_{\pi}\rightarrow 0$,
where 
the peculiar behavior of the NG-boson zero mode is derived: 
The NG-boson zero mode, when integrated over the LF, must behave as 
$\sim 1/m_{\pi}^2$. This ensures the non-vanishing matrix elements 
associated with the NG boson. However we encounter as an inevitable 
consequence that the LF charge is not conserved or even the current 
conservation breaks down even in the limit of $m_{\pi}\rightarrow
0$.\par 

Here we emphasize that {\it the NG theorem does not exist on the LF}. 
Instead we found
the singular behavior (\ref{omega}) which in fact {\it 
establishes existence of
the massless NG boson coupled to the current such that
$Q|0\rangle =0$ and $\dot{Q}\ne 0$}, quite analogously to
the NG theorem in the equal-time quantization which proves
existence of the massless NG boson coupled to the current such that
$Q^{\rm et}|0\rangle_{\rm et} \ne 0$ and $\dot{Q^{\rm et}}=0$ 
(opposite to the LF case!).
Thus the singular behavior of the NG-boson 
zero mode (\ref{omega}) (or (\ref{omega3})) may be understood as a 
remnant of the Lagrangian symmetry, an analogue of the NG 
theorem in the equal-time quantization. 

The zero mode problem was also discussed in the continuum theory with
careful
treatment of the boundary condition. 
It was demonstrated that as far as the sign function is
used for the commutator, the LF charge
does not annihilate the vacuum in sharp contrast to DLCQ, 
since the zero mode as an accumulating point
cannot be removed by simply dropping 
the exact zero mode with $p^+\equiv 0$
which is just measure zero. 
We also suggested that the ``$\nu$-theory'' 
might give a possible way out of this nontrivial vacuum problem 
in the continuum theory and give rise to the same result as that in
DLCQ.

The non-conservation of the SSB charge on the LF
was also stressed by Ida \cite{Ida} and Carlitz et al. \cite{CHKT} 
long time ago
in the continuum theory but not in DLCQ.
Their way to define the LF charge is somewhat similar to ours, namely,
the explicit mass of NG boson is  kept finite in order to
pick up the current matrix element with the NG-boson pole term dropped.
However, they discussed it in the continuum theory without consistent
treatment of the boundary condition and without realizing the zero mode
problem. If they were careful enough about
the boundary condition in the continuum theory, they would have
arrived at difficulty of the nontrivial vacuum 
as we mentioned before.
So it is essentially different from our argument in DLCQ.

Finally, we should mention that 
there is a
more serious zero-mode problem in the 
continuum 
LF theory, namely 
the no-go theorem found by Nakanishi and Yamawaki \cite{NY}.
The LF canonical commutator (\ref{FSf}) 
gives explicit expression of two-point Wightman 
function on LF: 
\begin{equation}
\left. \langle 0|\phi(x)\phi(0)|0\rangle \right|_{x^+=0}
= \frac{1}{2\pi}\int_{0}^{\infty}\frac{dp^+}{2p^+}e^{-ip^+x^-}\cdot 
\delta^{(2)}(x^{\bot}) \quad,
\label{WFLF}
\end{equation}
which is 
logarithmically divergent at $p^+ = 0$ and local
in $x^{\bot}$ and, more importantly, is
independent of the interaction and the mass. 
We can easily check this result in the free theory 
\cite{NYb} where the theory is 
explicitly solved in all space-time and the two-point 
Wightman function  
is given at any point $x$ by the well-known invariant delta function 
$\Delta^{(+)}(x;m^2)$ which is written in terms of the Hankel
function $K_1$ in the space-like region $x^2 <0$: 
\begin{eqnarray}
\langle 0|\phi(x)\phi(0)|0\rangle
&=&\Delta^{(+)}(x;m^2) = \frac{1}{(2\pi)^3}\int_{0}^{\infty}
\frac{dp^+ }{2p^+}\int_{-\infty}^{\infty} dp^{\perp} e^{-ip^-x^+
-ip^+x^- +ip^{\bot}x^{\bot}}
\nonumber\\
&=&\frac{m}{4\pi^2\sqrt{-x^2}}K_1 (m\sqrt{-x^2}) \quad\quad
(x^2<0)\quad.
\label{WFcov}
\end{eqnarray}
Restricting (\ref{WFcov}) to the LF, $x^+=0$, yields 
\begin{equation}
\left. \langle 0|\phi(x)\phi(0)|0\rangle \right|_{x^+=0}
=\frac{m}{4\pi^2\sqrt{x_{\bot}^2}}K_1 (m\sqrt{x_{\bot}^2})\quad,
\label{WFcov2}
\end{equation}
which is finite (positive definite), nonlocal in $x^{\bot}$ and
dependent on mass, in obvious contradiction to the above result
(\ref{WFLF}).
Hence, already for the free field the LF quantization fails to
reproduce the Lorentz-invariant theory.
Actually, the latter
Lorentz-invariant result (\ref{WFcov2}) is 
a consequence of the {\it mass-dependent} regularization of $1/p^+$ 
singularity 
at $p^+\rightarrow 0$
by the infinitely oscillating (mass-dependent) 
phase factor $e^{-ip^-x^+}=e^{-i(m^2+p_{\bot}^2)/2p^+ \cdot x^+}$ 
in the integral of (\ref{WFcov}) 
{\it before taking the LF restriction} $x^+ = 0$.
The LF quantization, restricting to $x^+=0$ beforehand, in fact
kills such a regularizing factor and leads to a wrong result
(\ref{WFLF}).  
Thus the LF restriction from the beginning 
loses all the information of dynamics 
carried  by the {\it zero mode as the accumulating point}.
This implies that {\it even a free theory does not exist on the LF}
\cite{NY}. 

One might suspect that this conclusion could be an artifact of too 
formal argument and
irrelevant to the actual physics, since one can construct free 
particle states, namely a free Fock space, with the correct spectra,
as far as the momentum space consideration is concerned. 
However, the above result implies that 
{\it quantum field on LF is ill-defined 
as the operator-valued distribution and so is the operator product on
LF.} Then it is rather difficult to construct a {\it realistic}
LF Hamiltonian (with interaction) in terms of
the products of local fields {\it on the same LF} 
in a way consistent with 
the Lorentz invariance, which would be a serious problem even for 
practical physicists. 

In fact, the above difficulty also applies to the interacting theory
satisfying 
the Wightman axioms (no-go theorem) \cite{NY},
in which case we have a spectral representation
(Umezawa-Kamefuchi-K\"allen-Lehmann representation)
for the commutator function: 
\begin{eqnarray}
\langle 0|[\phi(x), \phi(0)]|0\rangle 
&=&i \int_{0}^{\infty} d \mu^2 \rho (\mu^2) \Delta(x;\mu^2) 
\quad,
\nonumber\\
\int_{0}^{\infty} d \mu^2 \rho (\mu^2)&=&1, 
\quad \rho(\mu^2)\ge 0\quad.
\end{eqnarray}
If one assumed that LF restriction $x^+=0$ of 
the theory were well-defined, then it would follow:
\begin{eqnarray}
\langle 0|[\phi(x), \phi(0)]|0\rangle|_{x^+=0}=
i \int_{0}^{\infty} d \mu^2 \rho (\mu^2) \Delta(x;\mu^2)|_{x^+=0}
=-\frac{i}{4}\epsilon(x^-)
\delta^{(2)}(x^{\perp})   \quad,
\label{commfun}
\end{eqnarray}
since $i\Delta(x;\mu^2)|_{x^+=0}
=-\frac{i}{4}\epsilon(x^-)\delta^{(2)}(x^{\perp})
$ is independent of $\mu^2$. Taking 
the $p^+>0$ part of the Fourier component of the commutator function
 (\ref{commfun}), one would further obtain 
exactly the same result as (\ref{WFLF}) for the
two-point Wightman function at $x^+=0$, this time $\phi$ being 
the interacting Heisenberg field instead of the free field. 
On the other hand, the same
Wightman axioms yield the spectral representation also for 
the two-point Wightman function:
\begin{eqnarray}
\langle 0|\phi(x)\phi(0)|0\rangle
&=&\int_{0}^{\infty} d \mu^2 \rho (\mu^2) \Delta^{(+)}(x;\mu^2)\quad. 
\label{sprep}
\end{eqnarray}
Now, the LF restriction of (\ref{sprep}) depends on $\rho(\mu^2)$ and
does not agree with (\ref{WFLF}), since  
$\Delta^{(+)}(x;\mu^2)|_{x^+=0}$ given as (\ref{WFcov2}) does depend
on $\mu^2$, in sharp contrast to (\ref{WFLF})
which was  derived from 
 (\ref{commfun}).
Thus we have arrived at self-contradiction within the framework of 
Wightman axioms
under the assumption that LF restriction is well-defined.

An immediate way to resolve this trouble would be to define the theory
on the ``near LF'', $x^+\ne 0$, slightly away from the exact LF,
$x^+\equiv
0$, and then take 
the LF limit $x^+ \rightarrow 0$ only in the end of whole calculation 
as in (\ref{WFcov2}). 
In fact such a prescription was first proposed by Nakanishi and Yabuki
\cite{NYb}
in the continuum framework and later by 
Prokhvatilov et al. and others \cite{PF} in the
context of DLCQ.  
However, it was noted \cite{NY} 
that  
the price to pay in this approach is {\it non-vanishing vacuum
polarization} 
as in the equal-time quantization and 
hence {\it we must give up the trivial vacuum, or physical Fock space}, 
which is the most important feature of the LF quantization. Then there 
is no advantage of this approach over the equal-time quantization, 
concerning the simplicity of the vacuum in non-perturbative studies. 
Indeed, it was
demonstrated more explicitly \cite{PF2} that  the vacuum is 
nontrivial and there exists 
nontrivial renormalization 
in the LF Hamiltonian in this approach: It is no longer simple 
to solve dynamics compared with the equal-time quantization.

Thus, in spite of its difficulties with the above no-go theorem, 
we must take 
the quantization on the exact LF, $x^+\equiv 0$, from the beginning 
in order to keep
the trivial vacuum and physical Fock space. 
Actually, the no-go theorem implies that the LF restriction is 
not compatible with the Wightman axioms.  
Therefore, in order to make the   
{\it theory well-defined on the
exact LF,} 
we are forced to  
{\it give up some of the Wightman axioms, most naturally 
the Lorentz invariance. }  
Indeed, DLCQ defined on the exact LF is such a theory: The theory itself
explicitly violates the Lorentz invariance for $L<\infty$ 
and {\it never recovers it even in the limit of} 
$L \rightarrow \infty$ \cite{NY}, as we shall see later.
At the sacrifice of the Lorentz invariance,  
the trivial vacuum is in fact realized in DLCQ \cite{MY} as we have seen 
in this paper. The same is true in the $\nu$-theory \cite{NY} 
as we discussed in Section 5 and further demonstrate in the following.

In the $\nu$-theory
the two-point Wightman function for the free theory is given by
\cite{NY}:
\begin{eqnarray}
\Delta_{\nu}^{(+)} (x;m^2)
&=& \frac{c_\nu}{(2\pi)^3}\int_{0}^{\infty}
\frac{dp^+ }{2p^+} |p^+|^\nu \int_{-\infty}^{\infty} dp^{\perp}
e^{-ip^-x^+
-ip^+x^- +ip^{\bot}x^{\bot}},
\nonumber\\
&=&\frac{c_\nu [e^{i\pi/2}(x^+ - i0)]^\nu}{4\pi^2}
(\frac{m}{\rho})^{1+\nu}
K_{1+\nu} (m\rho) \nonumber\\
c_\nu (={\rm const.}) &>& 0 \quad(c_0=1),
\qquad \rho = [-2(x^+-i0)(x^- -i0)+x_\bot^2]^{\frac{1}{2}} \quad,
\label{WFnu}
\end{eqnarray}
where  the extra factor $c_\nu |p^+|^\nu$ is the regularization of the 
zero-mode singularity $1/p^+$ as was mentioned in Section 5. 
The previous non-commutativity 
between the integral of (\ref{WFcov}) and $x^+ \rightarrow 0$ 
is now traded for that between $\nu \rightarrow 0$ and 
$x^+ \rightarrow 0$.
If we take $\nu \rightarrow 0$ first and then $x^+\rightarrow 0$, 
we can reproduce correct
Lorentz-invariant result (\ref{WFcov2}), which is the same as
the procedure to take the ``near LF'' to the 
LF limit $x^+ \rightarrow 0$ \cite{NYb,PF,PF2}. 
If, on the other hand, 
we take $x^+ \rightarrow 0$ and then $\nu \rightarrow 0$, 
we arrive at the non-invariant answer (\ref{WFLF}) again.
Thus the theory itself (operator, Fock space, etc.)
violates Lorentz invariance 
and  
{\it never reproduces a 
Lorentz-invariant   
field theory even in the limit} $\nu\rightarrow 0$.
Conversely, the $\nu$-theory is well-defined on the exact LF
at the sacrifice of the Lorentz invariance (a part of Wightman axioms).

Now, the real problem is how to {\it recover Lorentz invariance of the
physical
quantity (c-number) like S matrix} which, unlike the Wightman function, 
{\it has no reference to the fixed LF}, 
{\it even though the theory itself, defined on the fixed exact LF,  
has no Lorentz-invariant limit}. 
Indeed,
it was pointed out \cite{NY} that
{\it as far as the perturbation theory is
concerned}, the {\it S matrix coincides} in the limit of $\nu\rightarrow
0$ 
{\it with the 
conventional Feynman rule result} which is Lorentz-invariant, 
{\it with one notable exception, 
namely the  
vanishing 
vacuum polarization graph} 
due to the modification of the zero-mode contribution. 
Note that $\nu\rightarrow 0$ is to be taken {\it after whole
calculation},
since the $\nu$-theory is defined on the exact LF only for $\nu>0$ 
(no $\nu=0$ theory exists on the exact LF, as dictated by the no-go
theorem). 
In fact, the Feynman  propagator of the $\nu$-theory takes the form
\cite{NY}:
\begin{eqnarray}
\Delta_{F,\nu} (x;m^2)
=\frac{ic_\nu }{(2\pi)^4} \int d \vec{p} |p^+|^\nu \int d p^-
\frac{e^{-ip^- x^+ +i\vec{p}\vec{x}} }{2p^-p^+ - p_{\bot}^2 -m^2+i0} 
\quad.  
\end{eqnarray}
Then the vacuum polarization graph  
calculated by the standard LF method \cite{CM} 
does vanish \cite{NY}:
\begin{eqnarray}
\int dp^+ |p^+|^\nu\int d p^- \frac {F(p^+p^-)}{2p^+p^- -m^2 +i 0}
&=&\int_{-\infty}^{\infty}d\lambda 
\tilde{F}(\lambda) \int_{0}^{\infty} d\xi 
\frac{e^ {-im^2\xi}}{i} \int dp^+|p^+|^\nu
\int dp^-  e^{i 2p^+p^-(\xi+\lambda)}\nonumber \\
&=& C
\int d p^+ |p^+|^\nu \delta(p^+) = 0 \quad,
\label{zerovp}
\end{eqnarray}
where 
$F$ and $\tilde{F}$ are a certain function and its Fourier transform
with respect to $\lambda$, respectively and  
$C= -\pi i \int d\lambda d\xi e^ {-im^2\xi} \tilde{F}(\lambda)/  
(\lambda +\xi) $
a numerical constant, and 
we have disregarded the transverse part which is irrelevant. 
Note that the {\it zero-mode contribution $\delta(p^+)$ 
has been modified by 
the extra factor $|p^+|^\nu$ ($\nu>0$) so as to yield zero vacuum 
polarization}.  
This is  consistent with
the previous argument in Section 5 
on the trivial vacuum $Q\vert 0\rangle=0$ in the
$\nu$-theory. 
In contrast, 
{\it all other graphs 
having no $\delta(p^+)$ would be 
unaffected by the extra factor $|p^+|^\nu$ and thus 
reproduce the usual Lorentz-invariant 
result in the $\nu\rightarrow 0$ limit}. 
The vanishing vacuum polarization (\ref{zerovp}) 
is in sharp contrast to the case where we 
take $\nu\rightarrow 0$ beforehand (no $|p^+|^\nu$ factor), 
 which actually corresponds to 
 the  prescription \cite{NYb,PF,PF2} approaching from 
``near LF'' to LF, with $\nu=0$: 
In such a case we have a non-vanishing
vacuum polarization 
as in the equal-time quantization, the whole contribution coming
from  
the zero mode $\int dp^+ \delta(p^+)\ne 0$, 
 as was noted by Chang and  Ma \cite{CM}.

We can also expect the same situation in DLCQ. The theory itself is
not Lorentz-invariant,
since the two-point Wightman function 
in the free theory
 takes the form:
\begin{eqnarray}
\Delta_{\rm DLCQ}^{(+)}(x;m^2)\vert_{x^+=0} &=&\frac{1}{2\pi}
\sum_{n>0} \frac{\pi}{L}\frac{1}{2p_n^+} e^{-ip_n^+ x^-}
\cdot \delta^{(2)} (x^\bot) \quad,\nonumber\\
p_n^+&=& \frac{n\pi}{L} \quad (n=1, 2, \cdot\cdot\cdot)\quad,
\end{eqnarray}
which coincides with (\ref{WFLF}) in the continuum 
limit of $L\rightarrow \infty$ (with 
$p_n^+=n\pi/L={\rm fixed}$),  
again in disagreement with the Lorentz invariance
\cite{NY}.
Note that the {\it sum does not include 
the zero mode 
} $n=0$, since the zero mode {\it in the free theory}
 vanishes  through 
the zero-mode constraint (\ref{zmconstraint2}). 
Although the {\it theory itself is not Lorentz-invariant}, 
we would  {\it reproduce the Lorentz-invariant result 
for the S matrix 
except for the vacuum polarization 
} in the continuum limit of $L\rightarrow \infty$,
{\it as far as the perturbation theory is concerned}. 
In fact, Feynman propagator in DLCQ takes the form:
\begin{eqnarray}
\Delta_{F,{\rm DLCQ}}(x;m^2)
&=&\frac{i}{(2\pi)^4}\sum_{n=\pm 1, \pm 2, \cdot\cdot\cdot}\frac{\pi}{L} 
\int d p^\bot d p^- 
\frac{e^{-ip^- x^+ -ip_n^+ x^- + i p^\bot x^\bot} }
{2p^-p_n^+ - p_{\bot}^2 -m^2+i0}\quad,
\end{eqnarray}
where again 
{\it the zero mode $n=0$ is not included in the sum}. 
When this is used in the 
Feynman rule for the perturbation, 
the {\it absence of the zero mode} $n=0$ actually
dictates that {\it vacuum polarization graph does vanish} similarly to
(\ref{zerovp}):
\begin{eqnarray}
\sum_{n=\pm 1, \pm 2, \cdot\cdot\cdot}\frac{\pi}{L}
\int d p^- \frac {F(p^+p^-)}{2p^+p^- -m^2 +i 0}
=C \sum_{n=\pm 1, \pm 2, \cdot\cdot\cdot} \frac{\pi}{L} \delta (p_n^+)
= 0\quad,
\label{zerovpdlcq}
\end{eqnarray} 
which is consistent with the trivial vacuum already established 
\cite{MY} through the zero-mode constraint (see Section 2).  
Note that the continuum limit ($\L\rightarrow \infty$) of  
(\ref{zerovpdlcq}) 
obviously disagree with the covariant result \cite{CM}
$
C\int d p^+ \delta(p^+) \quad (\ne 0). 
$
In contrast, {\it all other graphs 
having no $\delta(p_n^+)$ 
are insensitive to the zero mode $n=0$ and hence would 
coincide with the covariant result in such a limit}.

To summarize,  
the no-go theorem forbids the well-defined LF restriction of the 
Lorentz-invariant field theory 
due to the peculiarity of the zero mode
as an accumulating point in the continuum framework. 
Conversely, the theory  
defined on the exact LF such as DLCQ or $\nu$-theory,
although realizing the trivial vacuum and no vacuum polarization,  
would never recover the Lorentz-invariance 
even in the limit of $L\rightarrow \infty$ or $\nu\rightarrow 0$. 
Thus the Lorentz-invariant limit in such a theory 
can only be realized on the c-number 
physical quantity like S matrix which has no
reference to the fixed LF but not on the theory itself 
(Fock space, operator, etc). In fact, we have discussed that  
as far as the perturbation theory is concerned, both DLCQ and
$\nu$-theory 
would reproduce the Lorentz-invariant S matrix, while keeping the 
vacuum  polarization absent 
(no zero-mode contribution)
in accordance with the trivial vacuum. 
This was shown {\it through the explicit solution of the perturbative
dynamics} which is based on the interaction picture
with the propagator being given by the free theory
whose solution is completely 
known not only on a fixed LF $x^+=0$ but 
also on other region $x^+\ne 0$.

However, the real purpose of the LF quantization is 
to solve the dynamics {\it non-perturbatively} in a way much simpler
than the
equal-time quantization, based on 
the trivial vacuum and the physical Fock space for the interacting 
Heisenberg field. 
Then, in order to reproduce the Lorentz invariance 
without recourse to the perturbation theory, 
we actually would {\it need explicit solution of the non-perturbative
dynamics} itself, particularly the zero mode solution. 
Thus, recovering the Lorentz invariance is a {\it highly dynamical
issue} in the  LF quantization, the situation being 
somewhat analogous to the lattice gauge
theories. Then it 
remains a big challenge for the LF quantization to overcome 
the no-go theorem in 
the non-perturbative way. Particularly in DLCQ we  
would need to find  the
non-perturbative  
solution to the zero-mode constraint which 
might play a crucial role in taking 
the continuum limit $L\rightarrow \infty$ so as to recover  
the Lorentz invariance in the physical quantity (c-number).

In this paper we did not attempt to solve the above no-go
theorem in the non-perturbative sense.
Instead, in DLCQ (or in the $\nu$-theory), 
we only made a rather modest attempt
to solve the easier one, namely, to formulate the SSB in a manner
consistent
with the trivial LF vacuum. 
No doubt, fully non-perturbative solution to the no-go theorem is highly
desired.
Much work is to be done towards revealing the non-perturbative structure
of the
LF theory through the zero mode.

\noindent
{\large Acknowledgments}\par
We would like to thank K.G. Wilson for fruitful discussion.
We also appreciate stimulating discussion with
M. Taniguchi.
S.T. is supported by Max-Planck Gesellschaft Stipendium 
and thanks H.C. Pauli for encouragement and 
his interest in this work.
K.Y. is supported in part by the Grant-in-Aid of the Japanese 
Ministry of Education, Science and Culture (\# 08640365) 
and the Sumitomo Science Foundation. 
\newpage  

\appendix
\noindent


\appendix
\noindent

\section{DLCQ Canonical Commutator}\par
\my
Here we give a detailed derivation of 
the DLCQ canonical commutator (\ref{comm1}).

There are three constraints 
(\ref{Phi1}), (\ref{Phi2}) and (\ref{zmconstraint}):
\begin{eqnarray}
\Phi_{1}(x)
&\equiv& \pi_{\varphi}(x)-\partial_{-}
\varphi(x)\approx 0\quad, \\
\Phi_{2}(x)
&\equiv& \pi_{0}(x)\approx 0
\quad ,  \\
\Phi_3 (x) 
&\equiv& \frac{1}{2L}\int_{-L}^L dx^{-}
\left[(\mu^2-\partial_{\bot}^2)\phi+\frac{\partial V}{\partial
\phi}\right]  
\approx 0
\quad . 
\end{eqnarray}
Using (\ref{poisson2}) - (\ref{poisson4}), 
we compute
$C_{i,j}(x,y)\equiv\{\Phi_i(x), \Phi_j(y)\}$,
whose non-zero matrix elements are given by 
\begin{eqnarray}
C_{1,1}(x, y)
&=&(\partial^{y}_{-}-\partial^{x}_{-})\delta^{(3)}
(\vec{x}-\vec{y})\quad,\\
C_{1,3}(x, y)
&=&-\frac{1}{2L}\{\alpha(\vec{x})-\frac{1}{2L}\beta^{-1}(x^\bot)\}
\delta^{(2)} (x^{\bot}-y^{\bot})=-C_{3,1}(y,x)\quad,
\\
C_{2,3}(x, y)
&=&-\frac{1}{4L^2}\beta^{-1}(x^\bot)
\delta^{(2)} (x^{\bot}-y^{\bot})=-C_{3,2}(y,x)\quad,
\end{eqnarray}
where $x^+=y^+$ is understood and 
$\alpha (\vec{x})$ and $\beta^{-1}(x^{\bot})$ are given in 
(\ref{alphabeta}). 
Note that $\frac{1}{2L}\beta^{-1}$ is the
zero mode of $\alpha$.

Let us now calculate the inverse matrix $M_{i,j}(x,y)\equiv
(C^{-1})_{i,j}(x,y)$,
which is the essential step to obtain the Dirac bracket (\ref{DM}).  
It is easy to see that non-zero matrix elements 
of $M_{i,j}$ are given by 
\begin{eqnarray}
M_{1,1}(x, y)=
(C_{1,1})^{-1}(x, y)
&=&-\frac{1}{4}\left[\epsilon(x^--y^-)
-\frac{x^--y^-}{L}\right]\cdot\delta^{(2)}(x^{\bot}-y^{\bot})\quad,\\
M_{1,2}(x, y)&\equiv& p( \vec{x}, y^{\bot})\quad,\\
M_{2,1}(x, y)&\equiv& q(x^{\bot}, \vec{y})\quad,\\
M_{2,2}(x, y)&\equiv& r(x^{\bot}, y^{\bot})\quad,\\
M_{2,3}( x,  y )=
(C_{3,2})^{-1}( x, y) 
&=&4L^2
\beta(y^\bot)\cdot\delta^{(2)}(x^\bot-y^\bot)=-M_{3,2}(y,x)\quad,
\end{eqnarray}
where the functions $p$, $q$ and $r$ satisfy the equations: 
\begin{equation}
\int d\vec{y}C_{1,1}(x, y)p( \vec{y}, z^{\bot})+
\int dy^\bot C_{1,3}(x, y) M_{3,2}(y, z)=0 \quad, 
\end{equation}
\begin{equation}
\int d\vec{y} C_{3,1}(x,y)
M_{1,1}(y,z)+
\int dy^\bot C_{3,2}(x, y) q(y^{\bot}, \vec{z})=0 \quad, 
\end{equation}
\begin{equation}
\int d\vec{y}C_{3,1}(x,y) p( \vec{y}, z^{\bot})+
\int dy^\bot C_{3,2}(x,y)r(y^{\bot}, z^{\bot})=0 \quad. 
\end{equation}
We readily find the solutions:
\begin{eqnarray}
p(\vec{y}, z^{\bot})
&=& -\int d\vec{w}_1 dw_2^{\bot} 
 M_{1,1}(y, w_1)
 C_{1,3}(w_1, w_2) 
 M_{3,2}
 (w_2, z) \nonumber\\
&=& \frac{L}{2} \int_{-L}^{L}du^-\left[\epsilon(y^--u^-)-{y^--u^- 
 \over L}\right]\beta(z^\perp)
\alpha(u^-,z^{\bot})\delta^{(2)}(y^{\bot}-z^{\bot})
\quad,
\label{Mp} \\
q(y^{\bot}, \vec{z})
&=&-\int dw_1^{\bot} d\vec{w}_2 
 M_{2,3}(y, w_1) 
 C_{3,1}(w_1, w_2)
 M_{1,1}(w_2, z)
 \nonumber \\
&=&\frac{L}{2} \int_{-L}^{L}du^- \beta(y^\bot) 
 \alpha(u^-,y^{\bot})\left[\epsilon(u^--z^-)-{u^--z^- 
 \over L}\right]\delta^{(2)}(y^{\bot}-z^{\bot}) \nonumber \\
&=& -p(\vec{z},y^{\bot})
\quad,
\label{Mq}\\
r(y^{\bot}, z^{\bot})
&=& \int dw_1^{\bot} d\vec{w}_2 d\vec{w}_3 dw_4^{\bot}
 M_{2,3}(y, w_1) C_{3,1}(w_1, w_2) 
 M_{1,1}(w_2, w_3)\nonumber \\
&&\times C_{1,3}(w_3, w_4) 
 M_{3,2} (w_4, z)
\nonumber\\
&=& -L^2 \int dw_2^-dw_3^-\beta(y^\bot)\alpha(w_2^-, y^\bot)
\left[\epsilon(w_2^--w_3^-)-\frac{w_2^--w_3^-}{L}\right]
\beta(z^\bot) \alpha(w_3^-, z^\bot) \nonumber \\
&&\times \delta^{(2)}(y^\bot-z^\bot) \quad, 
\label{Mr}
\end{eqnarray}
where we have used 
\begin{equation}
\int_{-L}^{L} du^-  \left[\epsilon(u^--v^-)- \frac{u^--v^-}{L}\right] 
=\int_{-L}^{L} dv^- \left[\epsilon(u^--v^-)- \frac{u^--v^-}{L}\right]
=0\quad.      
\label{intsign}
\end{equation}

Now we are ready to derive the Dirac bracket for the full field 
$\phi=\varphi+\phi_0$:
\begin{eqnarray}
\{\phi(\vec{x}), \phi(\vec{y})\}_{DB} 
&=&\{\varphi(\vec{x}),\varphi(\vec{y})\}_{DB}
+\{\phi_0(x^{\perp}),\varphi(\vec{y})\}_{DB}
+\{\varphi(\vec{x}),\phi_0(y^{\bot})\}_{DB}\nonumber \\
&&+ \{\phi_0(x^{\bot}), \phi_0(y^{\bot})\}_{DB} \quad.
\label{DBfull}
\end{eqnarray}
According to (\ref{DM}), each Dirac bracket is evaluated as follows:
\begin{eqnarray}
\{\varphi(\vec{x}),\varphi(\vec{y})\}_{DB}
&=&
M_{1,1}(x,y)=-{1 \over 4}\left[\epsilon(x^--y^-)-{x^--y^- 
\over L}\right]\delta^{(2)} (x^{\bot}-y^{\bot})\quad,
\label{varphicomm}\\
\{\varphi(\vec{x}), \phi_0(y^{\bot})\}_{DB}
&=& \frac{1}{2L}\cdot
p(\vec{x}, y^{\bot})\nonumber \\
&=&\frac{1}{4}\int_{-L}^{L}du^-
\left[\epsilon(x^--u^-)-{x^--u^- \over L}\right]
\beta(y^{\bot}) \alpha(u^-,y^{\bot}) \nonumber\\
&&\times\delta^{(2)}(x^{\bot}-y^{\bot})\quad,
\label{var-0comm}\\
\{\phi_0(x^\perp),\varphi(\vec{y})\}_{DB}
&=& \frac{1}{2L}\cdot
q(x^\bot, \vec{y})= -\{\varphi(\vec{y}),\phi_0(x^{\bot})\}_{DB}
\nonumber \\
&=&\frac{1}{4}\int_{-L}^{L}du^-\beta(x^{\bot})
\alpha(u^-,x^{\bot})\left[\epsilon(u^--y^-)-{u^--y^- 
\over L}\right]\nonumber\\
&&\times\delta^{(2)}(x^{\bot}-y^{\bot})\quad,
\label{0-varcomm}\\
\{\phi_0(x^\bot), \phi_0(y^\bot)\}_{DB}
&=&\frac{1}{4L^2} \cdot r(x^{\bot},y^{\bot})\nonumber\\
&=&-\frac{1}{4}
\int du^-dv^- \beta(x^\bot)\alpha(u^-, x^\bot)
\left[\epsilon(u^--v^-)-\frac{u^--v^-}{L}\right] 
\beta(y^\bot) \alpha(v^-, y^\bot)\nonumber \\
&&\times 
\delta^{(2)}(x^\bot - y^\bot) \quad,
\label{0-0comm}
\end{eqnarray}
where use has been made of (\ref{Mp})-(\ref{Mr}) and 
the Poisson brackets (\ref{poisson2})-(\ref{poisson4}).

Combining (\ref{varphicomm})-(\ref{0-0comm}) into (\ref{DBfull}) and
making replacement $i\{A,B\}_{DB} \rightarrow [A,B]$, we arrive at 
the DLCQ canonical commutator (\ref{comm1}):
\begin{eqnarray}
\left[\phi(x),\phi(y)\right] 
&=&-\frac{i}{4}\int du^{-} dv^{-}\left[\beta(x^{\bot})
\alpha(u^-,x^{\bot})
                    -\delta(u^{-}-x^{-})\right] 
                      \left[\epsilon(u^--v^-)-
                      \frac{u^--v^-}{L} \right] \nonumber \\
&&\times\left[\beta(y^{\bot}) \alpha(v^-,y^{\bot}) -
\delta(v^{-}-y^{-})\right]
                   \delta^{(2)}(x^{\bot}-y^{\bot})\quad.
\end{eqnarray}
 
Next we demonstrate that the above commutator (\ref{comm1}) 
actually coincides with (\ref{comm2}), the one obtained in
Ref.\cite{MY},
up to operator ordering, i.e., in the sense of Dirac bracket.
First note that (\ref{0-0comm}) is actually vanishing and so is the 
zero-mode-zero-mode commutator $[\phi_0(x),\phi_0(y)]$:
\begin{equation}
r(x^{\bot},y^{\bot})=0 \quad,
\label{Mr0}
\end{equation}
where use has been made of (\ref{intsign}) and 
\begin{equation}
\int du^-\int dv^-\left[\epsilon(u^--v^-)- \frac{u^--v^-}{L}\right]
\alpha(u^-,x^{\bot})\alpha(v^-,y^{\bot})\delta^{(2)}(x^{\bot}-y^{\bot})=
0\quad.      
\end{equation}
Eq.(\ref{Mr0}) also follows from the antisymmetric property of 
$M_{i,j}(x,y)$ and the
one-dimensionality of $M_{2,2}$ in $x^-$ ($y^-$) direction: i,e,
$M_{2,2}(x,y)=r(x^{\bot},y^{\bot})=-M_{2,2}(y,x)=0$\quad. 
  
The sum of (\ref{var-0comm}) and (\ref{0-varcomm}) can be rewritten as  
\begin{eqnarray}
&&\frac{1}{4}\int du^{-} \beta(x^{\bot}) \alpha(u^-,x^{\bot})
                     \left[\epsilon(u^--y^-)-
                      \frac{u^--y^-}{L} \right]
                       \delta^{(2)}(x^{\bot}-y^{\bot})\nonumber\\
&&+ \frac{1}{4}\int du^{-} 
                      \left[\epsilon(x^--u^-)-
                      \frac{x^--u^-}{L} \right]
                  \beta(y^{\bot}) \alpha(u^-,y^{\bot})
                   \delta^{(2)} (x^{\bot}-y^{\bot}) 
                   \nonumber \\
&=&\frac{1}{4} \int
du^{-}\left[\epsilon(u^--y^-)+\epsilon(x^--u^-)\right]
\beta \alpha(u^-) \delta^{(2)} (x^{\bot}-y^{\bot}) \nonumber \\
&&- \frac{1}{4} \int du^{-}\left[\frac{u^--y^-}{L}+
\frac{x^--u^-}{L}\right]
\beta \alpha(u^-)  \delta^{(2)} (x^{\bot}-y^{\bot}) \nonumber \\
&=&\frac{1}{2} \int_{y^-}^{x^-} du^{-}\beta \alpha(u^-)
\delta^{(2)} (x^{\bot}-y^{\bot})
-\frac{1}{4}\left[\frac{x^--y^-}{L}\right]\delta^{(2)}
(x^{\bot}-y^{\bot})\quad, 
\end{eqnarray}
which is combined with (\ref{varphicomm}), yielding  
\begin{equation}
\left\{\phi(x),\phi(y)\right\}_{DB}=-\frac{1}{4}\left[
\epsilon(x^--y^-)-2 \beta \int_{y^-}^{x^-} dz^{-}\alpha(z^-)\right]\cdot
\delta^{(2)} (x^{\bot}-y^{\bot})\quad.
\end{equation}
Upon replacement $i\{A,B\}_{DB}\rightarrow [A,B]$, we finally 
obtain (\ref{comm2}).

\section{The Problem of Boundary Condition}\par
\my
We usually  assign  the boundary condition in DLCQ, 
because the ``space''coordinates are confined in the finite box 
$-L\leq x^- \leq L$. 
Besides such a practical reason, there is an inevitable reason 
why the boundary condition on $x^-$ direction  must be specified. 
This is not a specific problem of DLCQ, 
but a common problem to the whole LF quantization including the 
``continuum'' framework. 
To emphasize that the consistent LF quantization must be accompanied 
by the boundary condition we reexamine the special role of the 
boundary condition 
according to Steinhardt \cite{STH} and study  
what kind of boundary condition is consistent in  DLCQ.

Let us consider the ``continuum'' or 
``discretized'' LF quantization  of scalar model 
{\it without boundary condition} 
in the context of 
the Dirac quantization in Sect.2. 
Due to no boundary condition,
 the constraint for 
zero mode  will not appear. 
The only constraint appearing in the theory is 
\begin{equation}
\Phi(x)=\pi(x)-\partial_{-}\phi(x) \quad, 
\label{pc}
\end{equation}
whose Poisson bracket is given by 
\begin{equation}
\{\Phi(x), \Phi(y)\}=(\partial_{-}^y
-\partial_{-}^x)\delta ^{(3)}(\vec{x}-\vec{y})\quad. 
\end{equation}
Strictly speaking, we have infinitely many constraints which are  
expressible as linear combination of (\ref{pc}).

An important observation in \cite{STH} is 
that there is a subset of constraints which 
appears
 to be not only first 
class  but also second 
class. To see this, consider a linear combination of 
the primary constraint 
\begin{equation}
\Phi_0 \equiv \int dx^- \Phi(x) \quad , 
\end{equation}
which corresponds to the ``zero mode'' of $\Phi(x)$ 
in the discretized theory. 
Suppose that any surface term is neglected throughout the calculation, 
one  can  easily find   
\begin{equation}
\{\Phi_0, \Phi(x)\}=0 \quad . 
\end{equation}
This means that $\Phi_0$ is first class, because 
it should commute with any linear combination of $\Phi(x)$ 
as a consistency. 
However, this is not always the case, 
as the following example is illustrating:  
\begin{eqnarray}
\{\Phi_0, \int \epsilon(y^-)\Phi(y)dy^-\}&=&-2\int dx^- dy^-
\partial_{-}^y 
 \epsilon(y^-)\delta^{(3)}(\vec{x}-\vec{y})
\nonumber\\
 &=&-4 
 \delta^{(2)}(x^{\bot}-y^{\bot})\ne 0  \quad ,
\end{eqnarray}
where $\epsilon(x)$ is 
the sign function. This means that $\Phi_0$ is  second 
class in contradiction with the previous result. 
Actually,
  $\Phi_0$ is neither 
first class nor 
second class, which   
represents inconsistency hidden in the theory.   
This ambiguity reveals itself as the ambiguity 
of the inverse matrix of constraints, 
$C^{-1}$ in (\ref{DM}), and that of the Lagrange multiplier 
$v(x)$. It is easily shown that 
all such ambiguities can be removed,  
once the boundary condition at $x^-=\pm \infty$ or 
$x^-=\pm L$ is specified. 

Let us then study the possible boundary conditions in DLCQ. 
Although the same problem 
was studied by Steinhardt, he discussed it within 
the continuum framework and neglected all 
surface terms appearing  in the 
partial integrations. So we study the same problem by 
carefully treating surface terms in DLCQ. 
For this purpose,  we  generalize $\Phi_0$ and consider the 
following constraint which appears in the total Hamiltonian: 
\begin{equation}
\Phi[v]=\int_{-L}^{L}dx^-v(x)\Phi(x) \quad ,
\label{pc2}
\end{equation}
where $v(x)$ is a certain function (Lagrange multiplier) 
which satisfies the same boundary condition as $\phi(x)$ \cite{MY}. 
Once the boundary condition is specified,  
providing $\Phi[v]$ for all $v $ becomes  equivalent to 
providing $\Phi(x)$ for all $x$, which is nothing but the necessary
condition 
for consistency mentioned above. 
Moreover, we demand that 
the variation of canonical variable generated by (\ref{pc2}) 
must satisfy the same boundary condition. 
We can derive this condition by writing down the functional 
variation of $\Phi[v]$:  
\begin{eqnarray}
\delta \Phi[v] &=& \int_{-L}^L d x^- [v(x) \delta 
\pi(x)+{\partial}_{-}v(x) \delta \phi(x)] \nonumber\\   
&-& v(x^-=L)\delta \phi(x^-=L)+v(x^-=-L)\delta \phi(x^-=-L) \quad, 
\end{eqnarray}
where the first two terms on the r.h.s. give the canonical variation 
of the fields which preserve the same boundary condition as the
canonical 
variables.  
On the other hand,  the surface terms generally violate the boundary 
condition.  One can thus require the condition  
\begin{equation}
v(x^-=L)\delta\phi(x^-=L)=v(x^-=-L)\delta\phi(x^-=-L) \quad ,   
\label{gc}
\end{equation}
which is nothing but the discretized version of that derived in
\cite{STH}. 
This includes the periodic boundary condition studied in Ref.
\cite{MY}.  

Based on this condition  we investigate 
what kind of boundary condition can exist consistently. 
We pick up here some typical ones other than periodic case;

(I) the first boundary value: $\phi(x^-=L)=\phi(x^-=-L)=0$\quad,\par
(II) the second boundary value:  
$\frac{d}{dx^-}\phi(x^-=L)=\frac{d}{dx^-}\phi(x^-=-L)=0$\quad,\par
(III) the third boundary value: the mixed type of the above two 
conditions, \par
(IV) the anti-periodic boundary condition,\\ 
where the right hand sides of both (I) and (II) can be generalized to 
any value. Note that $\delta\phi(x)$ and $v(x)$ obey 
the same boundary condition as $\phi(x)$.  
Now, in the boundary condition (II) and (III) $\phi$ is left arbitrary 
at $x^-=\pm L$ and so are $\delta\phi(x)$ and $v(x)$,  and hence 
the boundary condition (II) and (III) 
do not generally satisfy the condition (\ref{gc}). 

Let us next consider the case (I), in which case it is 
helpful  to use mode expansion: 
\begin{equation}
\phi(x)=\sum_{n=1}^{\infty} a_n(x^+, x^{\bot}) \sin 
\{\frac{n\pi}{2L}(x^- +L)\} 
\quad. 
\end{equation}
The delta function should be interpreted as 
\begin{equation}
\delta(x^--y^-)=\frac{1}{L}\sum_{n=1}^{\infty}
\sin\{\frac{n\pi}{2L}(x^-+L)\}
\sin\{\frac{n\pi}{2L}(y^-+L)\}
\quad .
\end{equation}
It is shown that the inverse of the Dirac matrix, $C^{-1}(x, y)$,
satisfying the 
condition, 
$$
C^{-1}(x, y)=-C^{-1}(y, x), \quad C^{-1}(L, y)=-C^{-1}(-L, y)=0
$$ 
does not exist. Therefore the only constraint which may give the
consistent 
theory is the case of the anti-periodic boundary condition. 

\noindent
{\it Anti-periodic boundary condition} \cite {Mus}

The scalar field with the anti-periodic boundary condition 
is expanded by the complete sets 
$ \{\frac{1}{\sqrt{2L}} e^{-\frac{in \pi}{L}x^-} \}_{n=
\pm \frac{1}{2},\pm\frac{3}{2} \cdots} $, where there is 
no zero mode due to the anti-periodic boundary condition. 
For the Lagrangian (\ref{sl}), the only constraint is the 
primary constraint 
$\Phi(x)=\pi(x)-\partial_{-}\phi(x)$ which is second class. 
It is easily confirmed that the Poisson bracket 
$$
\{\Phi(x),\Phi(y)\}=(-\partial_{-}^{x}+\partial_{-}^{y})
\delta(x^{-}-y^{-})\delta^{(2)}(x^{\bot}-y^{\bot})
$$ 
has its inverse  
\begin{equation}
\{\Phi(x),\Phi(y)\}^{-1}=-{1 \over 4}\epsilon(x^--y^-)
\delta^{(2)}(x^{\bot}-y^{\bot})
\quad , 
\end{equation}
in the meaning of the delta function  
\begin{equation}
\delta(x^-)=\sum_{n}\frac{1}{2L}e^{-\frac{in\pi}{L}x^-} \quad 
\quad (n=\pm\frac{1}{2},\pm\frac{3}{2} \cdots) 
\quad . 
\end{equation}

Then the commutation relation is given by 
\begin{equation}
[\phi(x),\phi(y)]=-\frac{i}{4}\epsilon(x^--y^-)
\delta^{(2)}(x^{\bot}-y^{\bot})
\quad , 
\end{equation}
where both sides of the equation 
show the consistent 
behaviors at $x^{-}=\pm L$. 

Next we check the Poincar$\rm{\acute e}$ invariance of the theory. 
In the case of periodic boundary condition it is 
shown that the Poincar$\rm{\acute e}$ invariance is not 
recovered at least in the naive limit of $L \rightarrow \infty$ 
\cite{MY}. 
Hence it is interesting to study the same  problem for 
anti-periodic boundary condition. 

Let us first derive the equation of motion.  
The total Hamiltonian is described by 
\begin{equation}
H_T =\int d^3 \vec{x} \hspace{5pt}
[\frac{1}{2}\{(\partial_{\bot}\phi)^2 
+\mu^2 \phi^2\}+V(\phi)]+
\int d^3 \vec{x}\hspace{5pt}v(x)\Phi(x)
\quad , 
\end{equation}
where $v$ is the Lagrange multiplier. 

The consistency condition for the primary constraint $\Phi(x)$ reads  
\begin{eqnarray}
-2\partial_{-}v &=&(\mu^2-\partial^2_{\bot})\phi+
\frac{\partial V}{\partial \phi} 
\nonumber\\
&\equiv& \chi
\quad , 
\end{eqnarray}
where the surface term for $v(x)$ is dropped 
by the boundary condition. Combined with 
\begin{equation}
\partial_{+}\phi(x)=\{\phi(x), H_T \}=v(x) \quad, 
\end{equation} 
we obtain   
\begin{equation}
2\partial_+\partial_-\phi \approx -\chi
\quad . 
\end{equation}
By using this relation it is shown that 
\begin{eqnarray}
[\phi(x),P^{-}]&=&\int^L_{-L}d^3 {y} \{(\mu^2-\partial^2_{\bot})\phi(y)+
                 {\partial V \over \partial 
                 \phi}\}[\phi(x),\phi(y)], \nonumber \\
               &=&i\partial_+\phi(x)+{i \over 2}\{\partial_+\phi(x^-=L)
               \epsilon(x^--L)-\partial_{+}\phi(x^-=-L)
               \epsilon(x^-+L)\}, \nonumber \\
               &=&i\partial_+\phi(x)
\quad , 
\end{eqnarray}
where 
\begin{equation}
P^-=\int_{-L}^{L} d^3 x [\frac{1}{2}\{(\partial_{\bot}\phi)^2 
+\mu^2 \phi^2\}+V(\phi)]\quad . 
\end{equation}

Similar arguments can be applied to other Poincar$\rm{\acute e}$ 
generators: 
\begin{equation}
[ \phi(x), P^{\mu} ]=i{\partial}^{\mu}\phi(x)\quad ,
\end{equation}
\begin{equation}
[ \phi(x), M^{ij} ]=i(x^i {\partial}^j-x^j {\partial}^i)\phi(x)\quad ,
\end{equation}
\begin{equation}
[ \phi(x), M^{i+} ]=i(x^i {\partial}^{+}-x^+{\partial}^i)\phi(x)\quad, 
\end{equation}
\begin{equation}
[ \phi(x), M^{+-} ]=i(x^+ {\partial}^{-}-x^-{\partial}^+) 
\phi(x)\quad, 
\label{lorentz}
\end{equation}
which is compared with the periodic case \cite{MY} where (\ref{lorentz})
does not hold due to the zero mode. On the
other hand, we have
\begin{eqnarray}
[\phi(x),M^{-i}]&=&i(x^-\partial^{i}-x^i\partial^-)\phi(x) 
\nonumber\\
                &-&\frac{i}{4} [y^- \epsilon(x^--y^-)\partial_i \phi(y)]
                ^{y^-=L}_{y^-=-L}
\quad ,  
\end{eqnarray}
which is dependent on the box size $L$ and hence violates 
the Lorentz invariance even in the infinite volume limit 
as in the case of the periodic boundary condition \cite{MY}. 
Thus the Poincar$\rm{\acute e}$ invariance does 
not hold with respect to $M^{-i}$.  We need to devise 
an appropriate continuum limit instead of the naive limit 
to recover the  Poincar$\rm{\acute e}$  invariance of the theory. 
\newpage

\noindent
\section{Unbroken Phase of O(2)-Linear Sigma Model}\par
\my
We describe the unbroken phase of the O(2)-linear sigma model 
by treating the zero modes 
explicitly. 
In this phase we can explicitly check validity of our operator ordering, 
namely, the Weyl ordering used for studying the broken (SSB) phase 
of the sigma model in Sect.4.

For simplicity,  let us consider the bosonic part (\ref{blag}) 
with $\mu^2>0$ and $c\equiv 0$.
To solve the zero modes, it is convenient to 
divide them  into the classical and the operator parts 
as done in the Section 4. 
The solution of the zero-mode constraints for the 
classical part is  trivial, i.e. 
$v_{\pi}=v_{\sigma}=0$, implying  that 
the classical contribution to the vacuum expectation value is zero.      
On the other hand, the operator part of the zero-mode constraints is 
still too complicated to solve nonperturvatively. Then we solve them 
perturbatively again. 
The lowest order solutions of the operator part of the zero 
modes are derived as  
\begin{eqnarray}\label{symm}
\omega_{\pi}&=&\frac{\lambda}{2L}(-\mu^2
+\partial^2_{\bot})^{-1}\int_{-L}^{L}
dy^{-}(\varphi^3_{\pi}+\varphi_{\pi}\varphi_{\sigma}^2)\quad,\\
\omega_{\sigma}&=&\frac{\lambda}{2L}
(-\mu^2+\partial^2_{\bot})^{-1}\int_{-L}^{L}
dy^{-}(\varphi^3_{\sigma}+\varphi_{\sigma}\varphi_{\pi}^2)
\quad , 
\end{eqnarray}
both of which include no divergence and thus well-defined. 
These explicit solutions confirm the well-known properties in the 
unbroken phase. 

First of all, the same  transformation law as that in 
the equal-time quantization 
\begin{equation}
[Q, {\pi}]=i{\sigma} \quad, \quad \quad 
[Q, {\sigma}]=-i{\pi}\quad ,  
\label{tr}
\end{equation}
follows from the direct calculation 
\begin{equation}
[Q, \omega_{\pi}]=i\omega_{\sigma}\quad, \quad \quad 
[Q, \omega_{\sigma}]=-i\omega_{\pi}\quad ,  
\end{equation}
where 
\begin{equation}
Q=\int d^3\vec{x}
(\partial_{-}\varphi_{\sigma}\varphi_{\pi}-\partial_{-}
\varphi_{\pi}\varphi_{\sigma}) \quad .  
\end{equation}

Second, (\ref{tr}) is consistent to the trivial 
property of the LF
vacuum. Actually 
$\langle \omega_{\pi} \rangle =
\langle \omega_{\sigma} \rangle =0$ 
in addition to 
$\langle \varphi_{\pi} \rangle =\langle \varphi_{\sigma} \rangle =0$ 
lead to 
\begin{equation}
\langle[Q, \pi]\rangle= 
\langle[Q, \sigma]\rangle=0
\quad , 
\end{equation}
which is consistent with $Q\vert 0\rangle=0$. 

Now, let us check the operator ordering by checking the 
charge conservation which should hold 
in the unbroken phase in contrast to the broken phase. 
In fact an  arbitrary operator ordering would 
not necessarily lead to the charge conservation 
due to the existence of zero modes. 
One should find such an operator ordering as to satisfy the charge 
conservation and the {\it Weyl ordering} actually does it. 
Indeed it is easy to show 
\begin{equation}
i\frac{dQ}{dx^{+}}=[Q, H]=0
\quad ,
\end{equation}
under the Weyl ordering.    
It is this property that the  Weyl ordering was 
assumed even in the case of the broken phase in Sect.4. 
Although only the lowest order solution was explicitly 
examined, the same results stated above are valid beyond it. 

\newpage

\noindent
\section{The Structure of the Zero-Mode Constraint}\par
We reexamine the zero-mode constraint for the NG boson with the 
mass term. The crucial feature of the NG phase is the 
singular behavior of the zero mode (\ref{omega}) which is 
explicitly confirmed by  
the lowest order solution of the linear sigma model. 
The purpose of this Appendix is to confirm (\ref{omega})  
beyond the leading approximation. 
For simplicity, consider the bosonic sector of the linear sigma 
model in (\ref{op-zero2}) and (\ref{op-zero}). 
By redefining the lowest order solution 
\begin{eqnarray}
\omega_{\pi}^{(1)}
&=&\displaystyle{\frac{\lambda}{2L}\int_{-L}^{L}dx^-
(-m_{\pi}^2+\partial_{\bot}^2)^{-1}
(\varphi_{\pi}^3
+\varphi_{\pi}\varphi_{\sigma}^2+2v\varphi_{\pi}\varphi_{\sigma})}
\nonumber \\ 
&\equiv& {\cal D}_{\pi}\Phi_{\pi}(x^+, x^{\bot}) \quad ,
\end{eqnarray}
and
\begin{eqnarray}
\omega_{\sigma}^{(1)}
&=&\displaystyle{\frac{\lambda}{2L}\int_{-L}^{L}
dx^-(-m_{\sigma}^2+\partial_{\bot}^2)^{-1}(\varphi_{\sigma}^3
+\varphi_{\sigma}\varphi_{\pi}^2+v\varphi^2_{\pi}
+3v\varphi^2_{\sigma})}\nonumber \\ 
&\equiv& {\cal D}_{\sigma}\Phi_{\sigma}(x^+, x^{\bot}) \quad,
\end{eqnarray}
where ${\cal D}_{\pi}\equiv(-m_{\pi}^2+\partial_{\bot}^2)^{-1}$ and 
${\cal D}_{\sigma}\equiv(-m_{\sigma}^2+\partial_{\bot}^2)^{-1}$, 
the second order perturbative 
solution of the NG-boson zero mode is given by 
\begin{eqnarray}
\omega_{\pi}^{(2)}
&=&\displaystyle{
                  \frac{\lambda}{2L} \int_{-L}^{L} dx^-
                }
  {\cal D}_{\pi}
  \Bigl\{
   \frac 1{2}[({\cal D}_{\sigma}\Phi_{\sigma})
              \varphi_{\sigma}\varphi_{\pi}+
              \varphi_{\sigma}({\cal
D}_{\sigma}\Phi_{\sigma})\varphi_{\pi}
              +\varphi_{\pi}({\cal
D}_{\sigma}\Phi_{\sigma})\varphi_{\sigma}
              +\varphi_{\sigma}\varphi_{\pi}({\cal
D}_{\sigma}\Phi_{\sigma})
             ] \nonumber \\
   &+&[({\cal D}_{\pi}\Phi_{\pi})\varphi_{\pi}^2        
       +\varphi_{\pi}^2({\cal D}_{\pi}\Phi_{\pi})
       +\varphi_{\pi}({\cal D}_{\pi}\Phi_{\pi})\varphi_{\pi}
      ]
   +\frac1{2}[({\cal D}_{\pi}\Phi_{\pi})\varphi_{\sigma}^2
              +\varphi_{\sigma}^2({\cal D}_{\pi}\Phi_{\pi})
             ]
\Bigr\}
\quad,
\end{eqnarray} 
which leads to 
\begin{eqnarray}
\int d^3 \vec{x} \hspace{5pt}\omega_{\pi}^{(2)}&\sim& 
-\frac{\lambda}{m^2_{\pi}} \int d^3 \vec{x} 
\Bigl\{
\frac 1{2}[({\cal D}_{\sigma}\Phi_{\sigma})
\varphi_{\sigma}\varphi_{\pi}+
\varphi_{\sigma}({\cal D}_{\sigma}\Phi_{\sigma})\varphi_{\pi}
+\varphi_{\pi}({\cal D}_{\sigma}\Phi_{\sigma})\varphi_{\sigma}
+\varphi_{\sigma}\varphi_{\pi}({\cal D}_{\sigma}\Phi_{\sigma})] 
\nonumber \\
&+& [(\partial_{\bot}^{-2}\Phi_{\pi})\varphi_{\pi}^2    
+\varphi_{\pi}^2(\partial_{\bot}^{-2}\Phi_{\pi})
+\varphi_{\pi}(\partial_{\bot}^{-2}\Phi_{\pi})\varphi_{\pi}]
\nonumber \\
&+&\frac1{2}[(\partial_{\bot}^{-2}\Phi_{\pi})\varphi_{\sigma}^2
+\varphi_{\sigma}^2(\partial_{\bot}^{-2}\Phi_{\pi})] 
\Bigr\} 
\nonumber \\
&\sim& \frac{1}{m^2_{\pi}} \quad, \nonumber 
\end{eqnarray} 
in the  
$m_{\pi}\rightarrow 0$ limit. 
Repeating the similar procedure, such singular behaviors of the 
zero mode for higher order perturbative solutions ($n>1$), 
$$
\int d^3 \vec{x} \hspace{5pt}\omega_{\pi}^{(n)}\sim
\frac{1}{m^2_{\pi}} \quad, \nonumber 
$$
can also be confirmed inductively. 

Here it may be interesting to consider the same problem in 
two-dimensions 
where a continuous symmetry cannot be broken spontaneously 
 due to Coleman's theorem \cite{Cm}. In two 
dimensions, the 
leading order solution of the zero modes is given by 
\begin{eqnarray}
\omega_{\pi}^{(1)}
&=& -
\frac{1}{m_{\pi}^2}
\Phi_{\pi} \quad ,\\
\omega_{\sigma}^{(1)}
&=&-
\frac{1}{m_{\sigma}^2}
\Phi_{\sigma} \quad,
\end{eqnarray}
where $\Phi_{\pi, \sigma}$ depend on $x^+$ only. 
Without integral in $x^{\bot}$, 
$\omega_{\pi}^{(1)}$ {\it by itself} becomes divergent as 
$m_{\pi}\rightarrow 0$ in sharp contrast to the 
four-dimensional case. 
The peculiarity of two 
dimensions becomes clearer in the higher order 
solution. In fact, up to the operator 
ordering,  the second order solution 
\begin{equation}
\omega_{\pi}^{(2)}
=\displaystyle{\frac{\lambda}{2L}\int_{-L}^{L}
dx^-}\frac{1}{m_{\pi}^2}
\Bigl\{
\frac{1}{m_{\pi}^2}\Phi_{\pi}\varphi_{\sigma}^2
+\frac{2}{m_{\sigma}^2}\Phi_{\sigma}
+\frac{3}{m_{\pi}^2}\Phi_{\pi}\varphi_{\pi}^2
\Bigr\} \quad, 
\label{sec2}
\end{equation} 
leads to 
\begin{equation}
\omega_{\pi}^{(2)}\sim \frac{1}{m_{\pi}^4} 
\quad \quad (m_{\pi}\sim 0) \quad. 
\end{equation}
In general the $n$-th order perturbative solution of the 
NG-boson zero mode behaves as 
\begin{equation}
\omega_{\pi}^{(n)}\sim 
\frac{1}{m_{\pi}^{2n}} \quad.
\end{equation} 
The singular behavior becomes  worse in higher orders. 
In other words,  
the NG-boson field in two dimensions is ill-defined 
due to its wrong ``infrared behavior'' at $p^+\rightarrow 0$. 
In this sense  
the NG phase cannot be realized in two 
dimensions in conformity with 
Coleman's theorem \cite{Cm}, 
even if we apply our method by introducing the NG-boson mass 
as a regulator.

\end{document}